\documentstyle[12pt]{article}

\newcommand{\mysection}{\setcounter{equation}{0}\section}

\newcommand{\sqt}{{\sqrt{t}}}
\newcommand{\lnsqt}{{{\ln {t-1 \over \sqrt{t}}}}}
\newcommand{\sqrq}{{\sqrt{q^4-4 m^2 q^2}}}
\newcommand{\mpq}{{(m^2+q^2)}}
\newcommand{\mmq}{{(m^2-q^2)}}
\newcommand{\moq}{{\left({m^2 \over -q^2}\right)}}
\newcommand{\lam}{{\lambda}}
\newcommand{\lsq}{{\lambda^{1 \over 2}}}
\newcommand{\lcu}{{\lambda^{3 \over 2}}}
\newcommand{\lfi}{{\lambda^{5 \over 2}}}
\newcommand{\lse}{{\lambda^{7 \over 2}}}
\newcommand{\lnx}{{\ln\xi}}
\newcommand{\smm}{{(\hat s-m^2)}}
\newcommand{\smq}{{(\hat s-m^2-q^2)}}
\newcommand{\cvs}{{{\cal R}^{S+V}}}
\begin{document}
{\tt hep-ph/0102280}\\
\begin{flushright}
MRI-P-010204\\
\end{flushright}
\begin{flushright}
INLO-PUB-01/2001
\end{flushright}
\begin{flushleft}
\end{flushleft}
\vskip 1.0cm
\centerline{\large\bf {QCD and power corrections to sum rules}} 
\vskip 1.0cm
\centerline{\large\bf {in deep-inelastic lepton-nucleon scattering}}
\vskip 2.0cm
\centerline {\sc V. Ravindran}
\centerline{\it Harish-Chandra Research Institute,}
\centerline{\it Chhatnag Road, Jhusi, Allahabad-211019,}
\centerline{\it India.}
\vskip 1.0cm
\centerline {\sc W.L. van Neerven}
\centerline{\it Instituut-Lorentz,}
\centerline{\it University of Leiden,}
\centerline{\it PO Box 9506, 2300 RA Leiden,}
\centerline{\it The Netherlands.}
\vskip 1.0cm
\centerline{February 2001}
\vskip 2.0cm
\centerline{\bf Abstract}
\vskip 0.3cm
In this paper we study QCD and power corrections to sum rules which
show up in deep-inelastic lepton-hadron scattering. Furthermore we
will make a distinction between fundamental sum rules which can be
derived from quantum field theory and those which are of a phenomenological
origin. Using current algebra techniques the fundamental sum rules can be 
expressed into expectation values of (partially) conserved (axial-) vector 
currents sandwiched between hadronic states. These expectation values yield
the quantum numbers of the corresponding hadron which are determined by the 
underlying flavour group $SU(n)_F$. In this case one can show
that there exist an intimate relation between the appearance of power
and QCD corrections. The above features do not hold for 
phenomenological sum rules, hereafter called non-fundamental. They have
no foundation in quantum field theory and they mostly
depend on certain assumptions made for the structure functions like
superconvergence relations or the parton model. Therefore
only the fundamental sum rules provide us with a stringent test of QCD.\\[3mm] 

\mysection{Introduction}
Sum rules in deep-inelastic lepton hadron scattering provide us with one
of the most beautiful tools to test the predictions of perturbative
QCD \cite{neer}. This is because they can be expressed into integrals of the 
type
$\int_0^1dx~\Delta F^N(x,q^2)= A_N$ where $\Delta F^N(x,q^2)$ either denotes 
a structure function or a combination of structure functions and $N$ 
represents the corresponding hadron in the initial state of the deep-inelastic 
process. In this way one gets rid of the unknown $x$-dependence which is
due to non-perturbative effects. However this statement only holds if $A_N$
can be determined in an unambiguous way as we will elucidate in this paper.
Furthermore sum rules can be computed up to much higher orders in perturbation 
theory than other quantities which are mostly known up to next-to-leading 
order only. 
Examples are the Bjorken sum rules \cite{bjork1}, \cite{bjork2} and the Gross- 
Llewellyn Smith sum rule \cite{grll} which have been calculated up to order 
$\alpha_s^3$ in \cite{latk}, \cite{lave}. The same features are shown by
the total cross section in $e^+~e^- \rightarrow hadrons$ \cite{gkl} and the
width of the Z-boson \cite{lrs}.
The only problem is that it is very hard to measure the sum rules 
experimentally since the structure functions are only known for
a limited range of $x$. Therefore the uncertainties are due to
extrapolations of the structure functions into the small and large $x$-region. 
Fortunately the small $x$-region is not so important since all sum rules
only hold for the non-singlet parts of the structure functions which 
tend
to zero when $x \rightarrow 0$. The large $x$-region is important and here the
data are mainly coming from fixed target experiments. More information
about this region will come after the upgrade of HERA \cite{wagner}.
Furthermore we place our hope on polarized electron-proton colliders
\cite{roge}, \cite{ansel} which allows us to measure the parity violating
structure functions $g_i(x,q^2)$ ($i=3-5$) for which some interesting
sum rules can be derived. In the literature one can find many sum rules
(see e.g. \cite{adda}, \cite{llew}, \cite{blko1}, \cite{blko2}). However from a 
theoretical point of view they cannot be put on an equal footing. In
this paper we make a clear distinction between fundamental and non-fundamental
sum rules. In the former case the quantity $A_N$ is given by the expectation
value of a conserved current or partially conserved axial vector current
sandwiched between the hadronic state $N$. Furthermore the (axial-) vector
currents are put in the adjoint representation of the underlying flavour group 
given by $SU(n)_F$ where $n$ denotes the number of light flavours. These
sum rules can either be derived from equal time current algebra \cite{gell} 
or from light-cone current algebra \cite{frge} \cite{coja}. Examples are
the sum rules given by Adler \cite{adler}, by Bjorken \cite{bjork1} 
\cite{bjork2} and
by Gross and Llewellyn Smith \cite{grll}. The Ellis-Jaffe sum rule \cite{elja}
does not
belong to this class since the singlet axial-vector current is not conserved
due to the Adler-Bell-Jackiw anomaly \cite{abj}. Therefore it acquires an
infinite renormalization which induces scaling violating terms in the 
perturbation series for $A_N$ so that it becomes a non-fundamental sum rule.
To the latter class also belong all sum rules which cannot be derived from
current algebra. For instance the sum rules given by Gottfried 
\cite{gott}, Burkhardt and Cottingham \cite{buco} cannot be expressed into
expectation values of (axial-)vector current operators. Many more sum rules
of this type can be found in \cite{blko2}. 
In this paper we will give a complete list of fundamental sum rules using the 
techniques of current algebra. Further we investigate the perturbation series 
which also includes a study of the power corrections.
It turns out that there is an intimate relation between the appearance of 
non-zero higher order corrections and the presence of power
corrections which can be either due to higher twist operators or mass 
dependent terms of the type $m^2/q^2$ which vanish in the limit 
$-q^2 \rightarrow \infty$. Here one has to bear in mind that sum rules do not 
acquire target mass correction since the spin of the operators (here (axial-) 
vector currents) is smaller than two. Apparently this also holds
when the sum rule cannot be related to expectation values of (axial-) vector
currents like the one given by Burkhardt and Cottingham \cite{buco} (see the 
treatment of target mass corrections in \cite{bltk}). The paper is organized
as follows. In section 2 we rederive the Adler sum rule for the unpolarized 
structure function $F_2(x,q^2)$ from the equal time
current algebra using infinite frame techniques. One of the features
is that this sum rule neither receives QCD corrections nor power contributions.
It turns out that a similar sum rule exists for the polarized structure 
function $g_4(x,q^2)$ appearing in neutrino-nucleon scattering. 
In section 3 we show that the remaining sum rules
can be obtained from light-cone current algebra. Since the region outside
the light-cone is not taken into account these sum rules acquire
power corrections and the perturbation series receives non-zero contributions
beyond lowest order. In section 4 we study the power corrections by keeping
a non-zero mass in the calculation of the first moment of the coefficient
function. Here we show that when mass dependent terms appear one also 
encounters non-vanishing higher order corrections. At the end we discuss
some peculiar sum rules for which $A_N=0$ up to order $\alpha_s$ in
perturbation theory. One of them is the Burkhardt Cottingham sum rule 
\cite{buco}. In the derivation we assume that the underlying flavour symmetry
is given by the group $SU(3)$. The formulae for $SU(4)$ which are more
appropriate at large $-q^2$ are presented in Appendix A. The long expressions
for the partonic structure functions  needed for the computation of the
QCD corrections to the sum rules are given in Appendix B.


\mysection{Infinite momentum frame techniques}
In this chapter we use the infinite momentum frame technique to
derive a sum rule for the longitudinally polarized structure function
$g_4(x,q^2)$ which is the analogue of the Adler sum rule \cite{adler}
in unpolarized (anti-) neutrino-nucleon scattering. Before the derivation
we first give the definitions for the structure functions which emerge
in deep-inelastic lepton-hadron scattering 
\begin{eqnarray}
\label{eq:2.1}
l(k,\lambda) + N(p,s) \rightarrow l'(k') + 'X'\,,
\end{eqnarray}
where $l$ and $l'$ denote the incoming and outgoing leptons and $N$
represents the incoming hadron. The inclusive final hadronic state is
given by $'X'$. Furthermore we have indicated between the brackets 
the momenta ($k,k',p$) and spins ($\lambda,s$) of the particles.
In lowest order of the electroweak standard model the above reaction proceeds 
via the exchange of one of the vector bosons $\gamma, Z$ (neutral current
process) or $W^{\pm}$ (charged current process). Following the notations
in \cite{blko1}, \cite{blko2} the hadronic tensor $W^{\mu \nu}$
is defined by 
\begin{eqnarray}
\label{eq:2.2}
W^{\mu \nu}_{(V_1V_2)}(p,q,s)&=& {1 \over 4 \pi} \int d^4 z\,
e^{i\,q\cdot z}\,<N(p,s)| ~[J^{\mu~\dagger}_{V_1} (z) ,J^{\nu}_{V_2}(0) ] ~
|N(p,s) >\,.
\end{eqnarray}
Here $V_1,V_2$ refer to the intermediate vector bosons $V_i=\gamma,Z,W^{\pm}$.
which appear in reaction (\ref{eq:2.1}). Furthermore we have $p^2=m^2$,
$s^2=-1$ and $s\cdot p=0$. In the case $V_1\not =V_2$ we only consider
the tensor $W^{\mu \nu}_{(V_1V_2)}+W^{\mu \nu}_{(V_2V_1)}$.
Using Lorentz covariance and time-reversal invariance we can express the
hadronic tensor in terms of fourteen structure functions
\begin{eqnarray}
\label{eq:2.3}
W^{\mu \nu}_{(V_1V_2)}(p,q,s)&=&-\tilde g^{\mu \nu} F^{V_1V_2}_1(x,q^2)
 +{\tilde p^{\mu} \tilde p^{\nu} \over p\cdot q} F^{V_1V_2}_2(x,q^2)
+i \epsilon^{\mu \nu \alpha \beta}
{p_\alpha q_\beta \over 2\, p\cdot q} F^{V_1V_2}_3(x,q^2)
\nonumber \\[2ex]
&&  +{q^\mu q^\nu \over p \cdot q} F^{V_1V_2}_4(x,q^2)
    +{p^\mu q^\nu + p^\nu q^\mu \over 2\, p\cdot q} F^{V_1V_2}_5(x,q^2)
\nonumber \\[2ex]
&&   +i m \epsilon^{\mu \nu \alpha \beta}
   {q_\alpha s_\beta \over p\cdot q} g^{V_1V_2}_1(x,q^2)
   +i m \epsilon^{\mu \nu \alpha \beta}
    {q_\alpha (p\cdot q s_\beta -s\cdot q p_\beta) \over
           (p\cdot q)^2 } g^{V_1V_2}_2(x,q^2)
\nonumber \\[2ex]
&&  +{m \over p\cdot q} \left({\tilde p^\mu \tilde s^\nu +\tilde p^\nu 
\tilde s^\mu \over 2} -s\cdot q {\tilde p^\mu \tilde p^\nu \over (p\cdot q)} 
\right ) g^{V_1V_2}_3(x,q^2)
\nonumber \\[2ex]
&&  + m s\cdot q {\tilde p^\mu \tilde p^\nu \over (p \cdot q)^2 } 
g^{V_1V_2}_4(x,q^2) -m \tilde  g^{\mu \nu} {s \cdot q \over p \cdot q} 
g^{V_1V_2}_5(x,q^2)
\nonumber \\[2ex]
&&   +i~m \epsilon^{\mu \nu \alpha \beta} {p_\alpha s_\beta \over p\cdot q}
 g^{V_1V_2}_6(x,q^2) +m~s\cdot q~{q^\mu q^\nu \over (p \cdot q)^2} 
g^{V_1V_2}_7(x,q^2)
\nonumber \\[2ex]
&&  +m~s\cdot q~{p^\mu q^\nu + p^\nu q^\mu \over 2 (p\cdot q)^2} 
g^{V_1V_2}_8(x,q^2) +m~{s^\mu q^\nu + s^\nu q^\mu \over 2 p \cdot q} 
g^{V_1V_2}_9(x,q^2)\,,
\end{eqnarray}
with the following shorthand notations
\begin{eqnarray}
\label{eq:2.4}
\tilde g^{\mu \nu}=g^{\mu \nu}-{q^{\mu} q^{\nu} \over q^2} \,,\quad 
\tilde p^{\mu}=p^{\mu}-{p\cdot q \over q^2} q^{\mu} \,, \quad 
\tilde s^{\mu}=s^{\mu}-{s\cdot q \over q^2} q^{\mu}\,.
\end{eqnarray}
Instead of the set of structure functions above one can make other choices
(see e.g. \cite{ji}) as long as all structure functions are linearly 
independent.
The unpolarized $F^{V_1V_2}_i(x,q^2)$ $(i=1-5)$ and the polarized structure 
functions $g^{V_1V_2}_i(x,q^2)$ $(i=1-9)$ depend besides on the virtuality of 
the intermediate vector boson $-q^2>0$ and the Bjorken scaling variable 
$x=-q^2/2 p\cdot q$ also on the mass of the hadron. Contraction of the 
hadronic tensor with the leptonic tensor provides us with the cross section
which can be found in Eq. (11) of \cite{blko2}. When the lepton masses are
neglected the leptonic current is conserved so that the structure functions
$F_4,~F_5$ and $g_7,~g_8,~g_9$ drop out of the cross section. Notice that
the structure function $g_6$ also contributes if $\vec s \parallel \vec q$
which is often ignored in the literature. The electroweak currents 
are in general decomposed into a vector current $V_a^{\mu}$ and an axial-vector
current $A_a^{\mu}$ where a indicates that these currents belong to the
adjoint representation of the flavour group $SU(n)_F$. Notice that the
structure functions $F_i$ ($i=1-5$) and $g_1,~g_2,~g_6$ only get contributions
from vector vector and axial-vector axial-vector correlation functions whereas
$F_3,g_3,~g_4,~g_5$ and $g_i$ ($i=7-9$) are determined by vector-axial-vector
combinations.

According to the hypothesis of Gell-Mann in \cite{gell}, the zero components of
the vector and axial-vector currents
satisfy the following equal time commutation (ETC) algebra 
\begin{eqnarray}
\label{eq:2.5}
\left [V_a^0(0,\vec z),V_b^0(0,\vec y) \right ] &=&
\left [A_a^0(0,\vec z),A_b^0(0,\vec y) \right ]=
i\delta^{(3)}(\vec z -\vec y)\,f_{abc}V_c^0(z)\,,
\nonumber\\[2ex]
\left [V_a^0(0,\vec z),A_b^0(0,\vec y) \right ]&=&
\left [A_a^0(0,\vec z),V_b^0(0,\vec y) \right ]=
i\delta^{(3)}(\vec z -\vec y)\,f_{abc}A_c^0(z)\,.
\end{eqnarray}
In the expressions above $f_{abc}$ denote the structure constants of the 
Lie-algebra of $SU(n)_F$
which are defined by $[\lambda_a,\lambda_b]=2\,i\,f_{abc}\,\lambda_c$
where $\lambda_a$ are the generators of the Lie-algebra.
These relations are satisfied irrespective of the origin of the currents
so that it does not matter whether they are composed of fermionic or 
bosonic fields. We can now compute the integral
\begin{eqnarray}
\label{eq:2.6}
\int_{-\infty}^{\infty}d\,q_0\,W_{ab,VV}^{00}&=&\frac{1}{2}
\int d^3 \vec z \,e^{-i\vec q\cdot \vec z}\langle N(p,s) |[V_a^0(0,\vec x),
V_b^0(0,\vec 0)]|N(p,s) \rangle = 
\nonumber\\[2ex]
&& \frac{i}{2}f_{abc} \langle N(p,s) |V_c^0(0)| N(p,s) \rangle \,.
\end{eqnarray}
Similar results are obtained for
\begin{eqnarray}
\label{eq:2.7}
\int_{-\infty}^{\infty}d\,q_0\,W_{ab,AA}^{0\nu}&=&\frac{i}{2}f_{abc} 
\langle N(p,s) |V_c^{\nu}(0)| N(p,s) \rangle\,,
\\[2ex]
\label{eq:2.8}
\int_{-\infty}^{\infty}d\,q_0\,W_{ab,VA}^{0\nu}&=&\frac{i}{2}f_{abc}
\langle N(p,s) |A_c^{\nu}(0)| N(p,s) \rangle\,,
\\[2ex]
\label{eq:2.9}
\int_{-\infty}^{\infty}d\,q_0\,W_{ab,AV}^{0\nu}&=&\frac{i}{2}f_{abc}
\langle N(p,s) |A_c^{\nu}(0)| N(p,s) \rangle\,.
\end{eqnarray}
The expectation values of the vector and axial-vector currents are given
by
\begin{eqnarray}
\label{eq:2.10}
\langle N(p,s) |V_c^{\nu}(0)| N(p,s) \rangle=\Gamma_c^N\,p^{\nu} \,,\quad
\langle N(p,s) |A_c^{\nu}(0)| N(p,s) \rangle=m\,\Gamma_c^{5,N}\,s^{\nu} \,.
\end{eqnarray}
In order to compute the sum rules we can choose an infinite momentum frame 
where the momenta have the following components
\begin{eqnarray}
\label{eq:2.11}
p=\left (P+\frac{m^2}{2\,P},\vec 0_{\perp},P\right )\,, \quad 
q=\left (\frac{m\,\nu}{P},\vec q_{\perp},0 \right )\,.
\end{eqnarray}
so that $\nu=p\cdot q/m$ is satisfied. In this frame the 
longitudinal spin has large components only. Because $s \cdot p=0$, the
components of the longitudinal spin can be written as 
\begin{eqnarray}
\label{eq:2.12}
s=\left (\frac{P}{m},0_{\perp},\frac{P}{m}+\frac{m}{2P}\right ) \,.
\end{eqnarray}
In the limit $P \rightarrow \infty$ we obtain
\begin{eqnarray}
\label{eq:2.13}
\int_{-\infty}^{\infty}d\,q_0\,W_{ab,VV}^{00}=\int_{-\infty}^{\infty}d\,q_0\,
W_{ab,AA}^{00} = P \,\int_{-\infty}^{\infty}\frac{d~\nu}{\nu} \, F_{2,ab,VV}
(\nu,q^2)\,,
\nonumber\\[2ex]
\int_{-\infty}^{\infty}d\,q_0\,W_{ab,AV}^{00}=\int_{-\infty}^{\infty}d\,q_0\,
W_{ab,VA}^{00} = P \,\int_{-\infty}^{\infty}\frac{d~\nu}{\nu} \, g_{4,ab,AV}
(\nu,q^2)\,.
\end{eqnarray}
In this paper we will not discuss the justification of the infinite momentum 
frame technique which seems to us rather formal. For more details
we refer to \cite{adda}.
From Eqs. (\ref{eq:2.6})-(\ref{eq:2.10}), (\ref{eq:2.13}) we obtain the 
following results.
\begin{eqnarray}
\label{eq:2.14}
&& \int_{-1}^1\frac{d~x}{x} \, F_{2,ab,VV}(x,q^2)= 
\int_{-\infty}^{\infty}\frac{d~\nu}{\nu} \, F_{2,ab,VV}(\nu,q^2)=\frac{i}{2}
f_{abc}\Gamma_c^N\,,
\nonumber\\[2ex]
&& \int_{-1}^1\frac{d~x}{x} \, g_{4,ab,AV}(x,q^2)=
\int_{-\infty}^{\infty}\frac{d~\nu}{\nu} \, g_{4,ab,AV}(\nu,q^2)=\frac{i}{2}
f_{abc}\Gamma_c^{5,N}\,.
\end{eqnarray}
The formulae above still do not represent the sum rules since they have
to be converted into integrals over the physical region $0<x<1$. Since the
ETC algebra in Eq. (\ref{eq:2.5}) only involves the Lie-algebra structure 
constants  $f_{abc}$ the sum rules can be only derived
for charged current processes. Choosing three flavours i.e. $n=3$ the
charged currents admit the following representation
\begin{eqnarray}
\label{eq:2.15}
J_{\pm}^{\mu}(y)=\Big (V_{1\pm i2}^{\mu}(y)-A_{1\pm i2}^{\mu}(y) \Big ) 
\cos \theta_c
 +\Big (V_{4 \pm i5}^{\mu}(y)- A_{4 \pm i5}^{\mu}(y)\Big ) \sin \theta_c\,,
\end{eqnarray}
where $\theta_c$ denotes the Cabibbo angle. Further we have introduced
the following shorthand notations
\begin{eqnarray}
\label{eq:2.16}
J_{+}^{\mu}\equiv J_{W^+}^{\mu} \,,\quad J_{-}^{\mu}\equiv J_{W^-}^{\mu}
\,, \quad J_{a\pm ib}^{\mu}\equiv J_a^{\mu}\pm i\,J_b^{\mu}\,,
\end{eqnarray}
so that the corresponding structure tensor reads
\begin{eqnarray}
\label{eq:2.17}
W^{\mu \nu}_{\pm}&=& {1 \over 4 \pi} \int d^4 z\,
e^{i\,q\cdot z}\,\langle N(p,s) | ~[J^{\mu~\dagger}_{\pm} (z) ,
J^{\nu}_{\pm}(0) ] ~| N(p,s)\rangle\,.
\end{eqnarray}
Further we infer from Eq. (\ref{eq:2.15}) the property 
$J_{\pm}^{\mu~\dagger}= J_{\mp}^{\mu}$. Notice that the current 
$J_{+}^{\mu}$ appears in the process 
\begin{eqnarray}
\label{eq:2.18}
\nu + N \rightarrow l^- +'X'\,, \quad \mbox{or} \quad 
l^+ + N \rightarrow \bar \nu +'X'\,,
\end{eqnarray}
which involves the $W^+$ exchange whereas $J_{-}^{\mu}$ shows up in
\begin{eqnarray}
\label{eq:2.19}
\bar \nu + N \rightarrow l^+ +'X'\,, \quad \mbox{or} \quad  
l^- + N \rightarrow  \nu +'X'\,,
\end{eqnarray}
which proceeds via $W^-$ exchange.
Substitution of the charged current Eq. (\ref{eq:2.15}) into the 
commutator of currents Eq. (\ref{eq:2.17}) leads to the result
\begin{eqnarray}
\label{eq:2.20}
W^{\mu \nu}_{\pm}&=&\pm 4\,i\,\cos^2 \theta_c(W^{\mu \nu}_{12,VV}-
W^{\mu \nu}_{12,AV}) \pm 4\,i\,\sin^2 \theta_c(W^{\mu \nu}_{45,VV}
-W^{\mu \nu}_{45,AV}) 
\nonumber\\[2ex]
&& \mp 4\,i\,\sin \theta_c \,\cos \theta_c
(W^{\mu \nu}_{24,VV}-W^{\mu \nu}_{15,VV}
-W^{\mu \nu}_{24,AV}+W^{\mu \nu}_{15,AV})\,.
\end{eqnarray}
Since the $AA$-part is equal to the $VV$-part (see Eq. (\ref{eq:2.5})) 
we do not distinguish 
them anymore. The same also holds for the $VA$-part and the $AV$-part.
From translation invariance (see Eq. (\ref{eq:3.8})) and 
$J_{\pm}^{\mu~\dagger}= J_{\mp}^{\mu~}$ one can derive
\begin{eqnarray}
\label{eq:2.21}
\langle N(p,s) | ~[J^{\mu~\dagger}_{\pm} (z) ,J^{\nu}_{\pm}(0) ] ~| N(p,s)
\rangle
=-\langle N(p,s) | ~[J^{\mu~\dagger}_{\mp} (-z) ,J^{\nu}_{\mp}(0) ] ~
| N(p,s)\rangle\,,
\end{eqnarray}
Substitution of the equation above into Eq. (\ref{eq:2.2}) provides us with
the relation
\begin{eqnarray}
\label{eq:2.22}
W^{\mu \nu}_{\pm}(p,q,s)= -W^{\mu \nu}_{\mp}(p,-q,s)\,.
\end{eqnarray}
Hence we have
\begin{eqnarray}
\label{eq:2.23}
F_2^{\rm W^-N}(x,q^2)=F_2^{\rm W^+N}(-x,q^2) \,, \quad 
g_4^{\rm W^-N}(x,q^2)=g_4^{\rm W^+N}(-x,q^2)\,.
\end{eqnarray}
We can now derive the following sum rules for three flavours. From the 
$VV$-part we obtain the Adler sum rule
\begin{eqnarray}
\label{eq:2.24}
&& \int_{-1}^1 \frac{d~x}{x}\,F_2^{\rm W^-N}(x,q^2)=
\int_0^1 \frac{d~x}{x}\left (F_2^{\rm W^-N}(x,q^2) - F_2^{\rm W^+N}(x,q^2)
 \right )= 
\nonumber\\[2ex]
&& \hspace*{4cm}(2\,I_3^N+3\,Y^N)+\cos^2 \theta_c\,(2\,I_3^N-3\,Y^N)\,,
\end{eqnarray}
where the vector charges (isospin, hypercharge and baryon number) are given by
\begin{eqnarray}
\label{eq:2.25}
\Gamma_3^N=2\,I_3^N \,,\quad \Gamma_8^N=\sqrt 3\,Y^N \,,\quad \Gamma_0^N=3\,B^N
\,, \quad \Gamma_6^N=0\,,
\end{eqnarray}
with $Y=S+B$ where $S$ denotes the strangeness of the hadron $N$.\\
In the case of polarized scattering we have the following analogue for
the Adler sum rule which follows from the $VA$-part.
\begin{eqnarray}
\label{eq:2.26}
&& \int_{-1}^1  \frac{d~x}{x} g_4^{\rm W^-N}(x,q^2)=
\int_0^1  \frac{d~x}{x} \left (g_4^{\rm W^-N}(x,q^2) - g_4^{\rm W^+N}(x,q^2)
 \right )=
\nonumber\\[2ex]
&& \Big \{-2\,I_3^N(F+D)-(3\,F-D)\}+\cos^2 \theta_c\,
\{-2\,I_3^N(F+D)+(3\,F-D)\Big \}\,.
\nonumber\\
\end{eqnarray}
For the axial vector charges we have
\begin{eqnarray}
\label{eq:2.27}
\Gamma_3^{5,N}=2\,I_3^N (F+D) \,, \quad \Gamma_8^{5,N}=\frac{1}{\sqrt 3}\,
(3\,F-D) \,, \quad \Gamma_6^{5,N}=0\,,
\end{eqnarray}
with $g_A=F+D=1.254 \pm 0.006$ and $3\,F-D=0.68\pm0.04$. These values follow
from the fact that the hadronic axial-vector 
current is only partially conserved i.e. $\partial_{\mu}A_c^{\mu}\not =0$.
In the case of the proton (p) and the neutron (n) the quantum numbers above
are given by
\begin{eqnarray}
\label{eq:2.28}
I_3^p=\frac{1}{2} \,, \quad B^p=1 \,,\quad S^p=0 \,,
\nonumber\\[2ex]
I_3^n=-\frac{1}{2} \,, \quad B^n=1 \,, \quad S^n=0 \,.
\end{eqnarray}
The most important feature of the two sum rules above is that they are exact
and therefore model independent contrary to the sum rules derived in the
next section. This means that they also hold beyond QCD. For instance the
ETC relations in Eq. (\ref{eq:2.5}) also holds for bosonic currents. The
reason is that at the tip of the light cone $z-y=0$ all ETC relations 
take the same form irrespective of the origin (bosonic or fermionic) of the
currents. Therefore in QCD the two sum rules have to be obeyed which implies
that they do not receive any power corrections of the type $(1/q^2)^p$ which 
e.g. can be attributed to mass corrections or to higher twist contributions.

\mysection{Light cone current algebra}
In this section we derive the sum rules which hold 
when the currents are expressed into fermionic fields. This happens in
QCD where the fermions are represented by the quarks.
According to the hypothesis made in \cite{frge}, \cite{coja} 
(see also \cite{jack}) the (axial-) vector currents
satisfy the following light-cone algebra
\begin{eqnarray}
\label{eq:3.1}
\left [V_a^{\mu}(z),V_b^{\nu}(y) \right ] &
\mathop{\mbox{=}}\limits_{\vphantom{\frac{A}{A}} (z-y)^2 \rightarrow 0}&
\left [A_a^{\mu}(z),A_b^{\nu}(y) \right ]
\mathop{\mbox{=}}\limits_{\vphantom{\frac{A}{A}} (z-y)^2 \rightarrow 0}
-\frac{1}{2}\partial_{\lambda}\,\Delta(z-y)
\nonumber\\[2ex]
&& \left [ i\, s^{\mu\nu\lambda\sigma}f_{abc} \left \{V_{\sigma c}(z,y)+
V_{\sigma c}(y,z)\right \} \right.
\nonumber\\[2ex]
&& \left. + s^{\mu\nu\lambda\sigma}d_{abc} \left \{
V_{\sigma c}(z,y)-V_{\sigma c}(y,z)\right \} \right.
\nonumber\\[2ex]
&& \left. - \epsilon^{\mu\nu\lambda\sigma}f_{abc} \left \{
A_{\sigma c}(z,y)-A_{\sigma c}(y,z)\right \} \right.
\nonumber\\[2ex]
&& \left. + i\,\epsilon^{\mu\nu\lambda\sigma}d_{abc} \left \{
A_{\sigma c}(z,y)+A_{\sigma c}(y,z)\right \} \right ]\,,
\end{eqnarray}
\begin{eqnarray}
\label{eq:3.2}
\left [A_a^{\mu}(z),V_b^{\nu}(y) \right ] &
\mathop{\mbox{=}}\limits_{\vphantom{\frac{A}{A}} (z-y)^2 \rightarrow 0}&
\left [V_a^{\mu}(z),A_b^{\nu}(y) \right ] 
\mathop{\mbox{=}}\limits_{\vphantom{\frac{A}{A}} (z-y)^2 \rightarrow 0}
-\frac{1}{2} \partial_{\lambda}\,\Delta(z-y)
\nonumber\\[2ex]
&& \left [ i\, s^{\mu\nu\lambda\sigma}f_{abc} \left \{A_{\sigma c}(z,y)+
A_{\sigma c}(y,z)\right \} \right.
\nonumber\\[2ex]
&& \left. + s^{\mu\nu\lambda\sigma}d_{abc} \left \{
A_{\sigma c}(z,y)-A_{\sigma c}(y,z)\right \} \right.
\nonumber\\[2ex]
&& \left. - \epsilon^{\mu\nu\lambda\sigma}f_{abc} \left \{
V_{\sigma c}(z,y)-V_{\sigma c}(y,z)\right \} \right.
\nonumber\\[2ex]
&& \left. + i\, \epsilon^{\mu\nu\lambda\sigma}d_{abc} \left \{
V_{\sigma c}(z,y)+V_{\sigma c}(y,z)\right \} \right ]\,.
\end{eqnarray}
Here $V_{\sigma c}(z,y)$ and $A_{\sigma c}(z,y)$ represent the bilocal
vector and axial-vector currents respectively.
Furthermore one has only kept the most singular pieces on the right-hand side
which contribute to the light cone behaviour of the commutators. Therefore
sub-leading terms, originating from mass corrections or higher twist 
contributions, 
are neglected. Another feature is the appearance of the symmetric structure
constants $d_{abc}$ which are defined by $\{\lambda_a,\lambda_b\}=2\,d_{abc}
\,\lambda_c$ where $d_{ab0}=\frac{2}{n}\,\delta_{ab}$ (see below 
Eq. (\ref{eq:2.5})). The causal function $\Delta(z-y)$ appearing in the 
expressions above is given by
\begin{eqnarray}
\label{eq:3.3}
i\,\Delta(z-y)=\frac{1}{(2\pi)^3}\int d^4p\,\epsilon(p^0)\,\delta(p^2)
\,e^{-ip\cdot (z-y)}\,,
\end{eqnarray}
which has the properties
\begin{eqnarray}
\label{eq:3.4}
\Delta(z-y)=0 \quad \mbox{for}\quad (z-y)^2<0 \,, \quad \partial_{\lambda}^z
\Delta(z-y)=-g_{\lambda 0}\delta^{(3)}(\vec z - \vec y)\,.
\end{eqnarray}
The current commutator algebra in Eqs. (\ref{eq:3.1}), (\ref{eq:3.2}) is 
satisfied by the free field bilocal currents
\begin{eqnarray}
\label{eq:3.5}
V_a^{\mu}(z,y)=\bar \psi(z) \gamma^{\mu}\,\frac{\lambda_a}{2} \psi(y) \,,\quad
A_a^{\mu}(z,y)=\bar \psi(z) \gamma^{\mu}\gamma^5\,\frac{\lambda_a}{2} \psi(y) 
\,.
\end{eqnarray}
Hence the local hermitian currents are obtained from the bilocal ones via
the relations
\begin{eqnarray}
\label{eq:3.6}
V_a^{\mu}(z)\equiv V_a^{\mu}(z,z) \,, \quad \qquad \qquad
A_a^{\mu}(z)\equiv A_a^{\mu}(z,z)\,,
\end{eqnarray}
where $\psi(z)$ represents the mass-less free Dirac field. From the current 
algebra in Eqs. (\ref{eq:3.1}), (\ref{eq:3.2}) one can express the
structure functions in Eq. (\ref{eq:2.3}) into Fourier transforms of
the bilocal currents for $(z-y)^2=0$ provided one takes the Bjorken limit
given by $-q^2 \rightarrow \infty$ with $x=-q^2/2p\cdot q$ is fixed. The
procedure starts by sandwiching the commutators between the hadronic state 
$|N(p,s)\rangle$ so that the
expectation values of the bilocal currents can be written as
\begin{eqnarray}
\label{eq:3.7}
\langle N(p,s) \mid V_c^{\sigma}(z,0)+V_c^{\sigma}(0,z) \mid N(p,s) \rangle &=&
2\,p^{\sigma}\,V_c^1(z^2,z\cdot p)+2\,i\,z^{\sigma}\,V_c^2(z^2,z\cdot p)\,,
\nonumber\\[2ex]
\langle N(p,s) \mid V_c^{\sigma}(z,0)-V_c^{\sigma}(0,z) \mid N(p,s) \rangle &=&
2\,p^{\sigma}\,\bar V_c^1(z^2,z\cdot p)+2\,i\,z^{\sigma}\,
\bar V_c^2(z^2,z\cdot p)\,,
\nonumber\\[2ex]
\langle N(p,s) \mid A_c^{\sigma}(z,0)+A_c^{\sigma}(0,z) \mid N(p,s) \rangle &=&
2\,s^{\sigma}\,m\,A_c^1(z^2,z\cdot p)
\nonumber\\[2ex]
&& -2\,i\,p^{\sigma}\,z \cdot s\, m\,A_c^2(z^2,z\cdot p)
\nonumber\\[2ex]
&& +2\,z^{\sigma}\,z \cdot s\,m\,A_c^3(z^2,z\cdot p)\,,
\nonumber\\[2ex]
\langle N(p,s) \mid A_c^{\sigma}(z,0)-A_c^{\sigma}(0,z) \mid N(p,s) \rangle &=&
2\,s^{\sigma}\,m\,\bar A_c^1(z^2,z\cdot p)
\nonumber\\[2ex]
&& -2\,i\,p^{\sigma}\,z \cdot s\, m\,\bar A_c^2(z^2,z\cdot p)
\nonumber\\[2ex]
&& +2\,z^{\sigma}\,z \cdot s\,M\,\bar A_c^3(z^2,z\cdot p)\,.
\end{eqnarray}
Under translation invariance the bilocal currents transform like
\begin{eqnarray}
\label{eq:3.8}
e^{-i\,\hat P\cdot z}\,J_c^{\mu}(0,z)\,e^{i\,\hat P\cdot z}=J_c^{\mu}(-z,0)\,,
\qquad J=V,A\,,
\end{eqnarray}
from which one can derive the following relations
\begin{eqnarray}
\label{eq:3.9}
V_c^1(z^2,-z\cdot p)&=&V_c^1(z^2,z\cdot p) 
\,, \quad V_c^2(z^2,-z\cdot p)= -V_c^2(z^2,z\cdot p)\,,
\nonumber\\[2ex]
\bar V_c^1(z^2,-z\cdot p)&=&- \bar V_c^1(z^2,z\cdot p) 
\,, \quad \bar V_c^2(z^2,-z\cdot p)= \bar V_c^2(z^2,z\cdot p)\,,
\nonumber\\[2ex]
A_c^k(z^2,-z\cdot p)&=&A_c^k(z^2,z\cdot p) \,, \quad k=1,3
\qquad A_c^2(z^2,-z\cdot p)= -A_c^2(z^2,z\cdot p)\,,
\nonumber\\[2ex]
\bar A_c^k(z^2,-z\cdot p)&=&- \bar A_c^k(z^2,z\cdot p)\,, \quad k=1,3
\qquad \bar A_c^2(z^2,-z\cdot p)= \bar A_c^2(z^2,z\cdot p)\,,
\nonumber\\
\end{eqnarray}
After insertion of the commutator algebra Eqs. (\ref{eq:3.1}), (\ref{eq:3.2}) 
into expression (\ref{eq:2.2}) one can compute the structure tensors which 
satisfy the relations $W^{\mu \nu,VV}_{ab}=W^{\mu \nu,AA}_{ab}$
and $W^{\mu \nu,AV}_{ab}=W^{\mu \nu,VA}_{ab}$. 
Here the superscripts $VV$, 
$AA$, $AV$, $VA$ indicate the type of currents which appear in the commutators 
above. A straightforward calculation provides us with
the following structure functions 
\begin{eqnarray}
\label{eq:3.10}
2\,x\,F_{1,ab}^{VV}(x,q^2)&=&F_{2,ab}^{VV}(x,q^2)=\frac{x}{2}\, \Big (i\,f_{abc}
\,V_c^1(x) +d_{abc}\,\bar V_c^1(x) \Big )\,,
\nonumber\\[2ex]
F_{3,ab}^{AV}(x,q^2)&=&-\frac{1}{2}\,\Big (i\,f_{abc}\,\bar V_c^1(x)
+d_{abc}\,V_c^1(x) \Big )\,,
\nonumber\\[2ex]
F_{4,ab}^{VV}(x,q^2)&=&F_{5,ab}^{VV}(x,q^2)=0\,,
\nonumber\\[2ex]
g_{1,ab}^{VV}(x,q^2)&=&\frac{i}{4}\,f_{abc}\Big (\bar A_c^1(x)+
\frac{\partial \bar A_c^2(x)}{\partial x} \Big )+\frac{1}{4}\,
d_{abc}\Big ( A_c^1(x) +\frac{\partial A_c^2(x)}{\partial x} \Big )\,,
\nonumber\\[2ex]
g_{2,ab}^{VV}(x,q^2)&=&-\frac{i}{4}\, f_{abc}\frac{\partial \bar A_c^2(x)}
{\partial x} -\frac{1}{4}\, d_{abc} \frac{\partial A_c^2(x)}{\partial x} \,,
\nonumber\\[2ex]
g_{3,ab}^{AV}(x,q^2)&=&\frac{i}{2}\,\,f_{abc} \Big (x\,A_c^1(x)-A_c^2(x) \Big )
+\frac{1}{2}\,\,d_{abc} \Big (x\,\bar A_c^1(x)-\bar A_c^2(x) \Big )\,,
\nonumber\\[2ex]
2\,x\,g_{5,ab}^{AV}(x,q^2)&=&g_{4,ab}^{AV}(x,q^2)=
\nonumber\\[2ex]
&& i\,\frac{x}{2}\, \,f_{abc}\Big (A_c^1(x)
+\frac{\partial A_c^2(x)}{\partial x} \Big )+\frac{x}{2}\,\,d_{abc}\Big (
\bar A_c^1(x) + \frac{\partial \bar A_c^2(x)}{\partial x} \Big )\,,
\nonumber\\[2ex]
g_{6,ab}^{VV}(x,q^2)&=&-\frac{i}{4}\,\,f_{abc}\Big (x\,\bar A_c^1(x)-
\bar A_c^2(x)\Big )-\frac{1}{4}\,\,d_{abc}\Big (x\,A_c^1(x)- A_c^2(x)\Big )\,,
\nonumber\\[2ex]
g_{7,ab}^{AV}(x,q^2)&=&0\,,
\nonumber\\[2ex]
x\,g_{9,ab}^{AV}(x,q^2)&=&- x\,g_{8,ab}^{AV}(x,q^2)=
\nonumber\\[2ex]
&& \frac{i}{4}\,f_{abc}\Big (x\,A_c^1(x)+ A_c^2(x) \Big )+\frac{1}{4}
\,d_{abc}\Big (x\, \bar A_c^1(x) +  \bar A_c^2(x) \Big )\,,
\end{eqnarray}
where we have defined
\begin{eqnarray}
\label{eq:3.11}
J_c^k(x)&=&\frac{1}{2\pi}\int_{-\infty}^{\infty}dz\cdot p\,\,
e^{-i\,x\,p\cdot z} \,J_c^k(0,p\cdot z) \,,\quad x=-\frac{q^2}{2p\cdot q}\,,
\nonumber\\[2ex]
J_c^k(0,p\cdot z)&=&\int_{-1}^1dx\,\,e^{i\,x\,p\cdot z}\,J_c^k(x) \,, \quad  
J=V,\bar V,A,\bar A\,.
\end{eqnarray}
Because of Eqs. (\ref{eq:3.1}), (\ref{eq:3.2}) one can identify 
$F_{i,ab}^{VV}=F_{i,ab}^{AA}$ and $F_{i,ab}^{AV}=F_{i,ab}^{VA}$. The same 
relations hold for the polarized structure functions $g_{i,ab}$. As we will
see later on these relations are also preserved in lowest order perturbation
theory except for $g_6$ which turns out to depend on the mass assignment
of the quarks in the Born reaction.  
In the above relations we have neglected all sub-leading terms which vanish
in the Bjorken limit $-q^2 \rightarrow \infty$ like $1/q^2$. In this
limit $F_4,~F_5,~g_7$ vanish and the matrix elements $V_c^2$, $\bar V_c^2$, 
$A_c^3$, $\bar A_c^3$ do not appear in the structure functions. 
Another feature is that the currents are composed of fermionic (spin half) 
fields (see Eq. (\ref{eq:3.5})) which
leads to the Callan-Gross relations $2xF_1=F_2$ \cite{cagr} and 
$2xg_5=g_4$ \cite{dicus} in Eq. (\ref{eq:3.10}). Furthermore one infers from 
Eq.  (\ref{eq:3.10}) that the unpolarized structure functions 
$F_i$ ($i=1,2,4,5$)
and the polarized structure functions $g_1,~g_2,~g_6$ only receive 
contributions from the commutators $[V,V]$ and $[A,A]$ (parity conserving)
whereas the quantities $F_3$ and $g_i$ ($i=3,4,5,7,8,9$) are
determined by the commutators $[A,V]$ and $[V,A]$ (parity violating) only. 
Further in the Bjorken limit only (leading) twist two contributions survive  
in the expressions for the structure functions $F_i$ ($i=1-3$) and $g_i$ 
($i=1,4,5$) whereas the structure functions $g_2,~g_3,~g_6$. 
also receive twist three contributions. For an analysis see e.g. \cite{blko1}, 
\cite{blko2}. We did a similar analysis for $g_8,~g_9$ and found that these
quantities are of twist three only. Notice that the latter 
structure functions do not show up in the cross section (see below 
Eq. (\ref{eq:2.4})) when the masses of the leptons can be neglected. 
Hence from Eq. (\ref{eq:3.10}) one can conclude
that the following matrix elements are of twist two
\begin{eqnarray}
\label{eq:3.12}
&& V_c^1(x)\,, \qquad A_c^1(x) +\frac{\partial A_c^2(x)}{\partial x}\,,
\nonumber\\[2ex]
&& \bar V_c^1(x)\,, \qquad \bar A_c^1(x)+\frac{\partial \bar A_c^2(x)}
{\partial x}\,.
\end{eqnarray}
For the moment we only limit ourselves to the structure functions which 
receive contributions of twist two only and postpone the discussion of the 
other ones to the end of this section.
From Eq. (\ref{eq:3.9}) one can derive the properties
\begin{eqnarray}
\label{eq:3.13}
V_c^1(-x)&=&V_c^1(x)
\,, \qquad V_c^2(-x)= -V_c^2(x)\,,
\nonumber\\[2ex]
\bar V_c^1(-x)&=&- \bar V_c^1(x)\,,
\qquad \bar V_c^2(-x)= \bar V_c^2(x)\,,
\nonumber\\[2ex]
A_c^k(-x)&=&A_c^k(x)\,, \quad k=1,3\,,
\qquad A_c^2(-x)= -A_c^2(x)\,,
\nonumber\\[2ex]
\bar A_c^k(-x)&=&- \bar A_c^k(x)\,, \quad k=1,3\,,
\qquad \bar A_c^2(z^2,-x)= \bar A_c^2(x)\,.
\end{eqnarray}
Like in the previous section we choose three flavours for our computations
so that one gets the representation for the charged current in
Eq. (\ref{eq:2.15}). For this choice the structure tensor in 
Eq. (\ref{eq:2.17}) becomes equal to
\begin{eqnarray}
\label{eq:3.14}
W^{\mu \nu}_{\pm}&=&2\,\cos^2 \theta_c\Big (W^{\mu \nu}_{11,VV}+
\,W^{\mu \nu}_{22,VV}+ 2\,i\,W^{\mu \nu}_{[12],VV}-W^{\mu \nu}_{11,AV}-
W^{\mu \nu}_{22,AV} 
\nonumber\\[2ex]
&& \mp 2\,i\, W^{\mu \nu}_{[12],AV}\Big ) +2\,\sin^2 \theta_c\Big (
W^{\mu \nu}_{44,VV}+W^{\mu \nu}_{55,VV}\pm 2\,i\,W^{\mu \nu}_{[45],VV}
-W^{\mu \nu}_{44,AV}
\nonumber\\[2ex]
&& -W^{\mu \nu}_{55,AV}\mp 2\,i\,W^{\mu \nu}_{[45],AV}\Big )
+ 4\,\sin \theta_c \,\cos \theta_c (W^{\mu \nu}_{\{14\},VV}+
W^{\mu \nu}_{\{25\},VV} \mp i\,W^{\mu \nu}_{[24],VV}
\nonumber\\[2ex]
&& \pm i\,W^{\mu \nu}_{[15],VV} -W^{\mu \nu}_{\{14\},AV} 
-W^{\mu \nu}_{\{25\},AV}\pm i\,W^{\mu \nu}_{[24],AV} 
\mp i\, W^{\mu \nu}_{[15],AV}\Big )\,,
\end{eqnarray}
with the definitions
\begin{eqnarray}
\label{eq:3.15}
W^{\mu \nu}_{\{ab\},VV}&=&\frac{1}{2}\Big (W^{\mu \nu}_{ab,VV}
+W^{\mu \nu}_{ba,VV}\Big )\,,
\nonumber\\[2ex]
W^{\mu \nu}_{[ab],VV}&=&\frac{1}{2}\Big (W^{\mu \nu}_{ab,VV}
-W^{\mu \nu}_{ba,VV}\Big )\,.
\end{eqnarray}
From the equations above one can derive the charged current structure functions
which are equal to
\begin{eqnarray}
\label{eq:3.16}
F_2^{\rm W^+ N}(x)/x&=&\cos^2 \theta_c \Big [\frac{4}{3}\,\bar V_0^1(x)
+\frac{2}{3}\sqrt 3\,\bar V_8^1(x) -2\,V_3^1(x) \Big ]
+ \sin^2 \theta_c \Big [\frac{4}{3}\,\bar V_0^1(x)
\nonumber\\[2ex]
&& +\bar V_3^1(x)-\frac{1}{3}\sqrt 3\,\bar V_8^1(x)-V_3^1(x)-\sqrt 3\,V_8^1(x) 
\Big ] 
\nonumber\\[2ex]
&& +2\,\sin \theta_c \,\cos \theta_c\,\Big [\bar V_6^1(x)+V_6^1(x)\Big ]\,.
\end{eqnarray}
The structure function $F_2^{\rm W^- N}$ is obtained from $F_2^{\rm W^+ N}$
by $V_a^1 \rightarrow - V_a^1$.
\begin{eqnarray}
\label{eq:3.17}
F_3^{\rm W^+ N}(x)&=&\cos^2 \theta_c \Big [\frac{4}{3}\, V_0^1(x)
+\frac{2}{3}\sqrt 3\, V_8^1(x) -2\,\bar V_3^1(x) \Big ]
+ \sin^2 \theta_c \Big [\frac{4}{3}\, V_0^1(x)
\nonumber\\[2ex]
&& + V_3^1(x)-\frac{1}{3}\sqrt 3\,V_8^1(x)-\bar V_3^1(x)-\sqrt 3\,\bar V_8^1(x)
\Big ]
\nonumber\\[2ex]
&& +2\,\sin \theta_c \,\cos \theta_c\,\Big [ V_6^1(x)+\bar V_6^1(x)\Big ]\,.
\end{eqnarray}
The structure function $F_3^{\rm W^- N}$ is obtained from $F_3^{\rm W^+ N}$
by $\bar V_a^1 \rightarrow - \bar V_a^1$.
\begin{eqnarray}
\label{eq:3.18}
g_1^{\rm W^+ N}(x)&=&\cos^2 \theta_c \Big [\frac{2}{3}\, \left (A_0^1(x)
+\frac{\partial A_0^2(x)}{\partial x} \right )
+\frac{1}{3}\sqrt 3\,\left (A_8^1(x)+\frac{\partial A_8^2(x)}{\partial x} 
\right )
\nonumber\\[2ex]
&& -\left (\bar A_3^1(x)+\frac{\partial \bar A_3^2(x)}{\partial x} 
\right ) \Big ] + \sin^2 \theta_c \Big [\frac{2}{3}\,\left (A_0^1(x)
+\frac{\partial A_0^2(x)}{\partial x} \right )
\nonumber\\[2ex]
&& +\frac{1}{2} \left (
A_3^1(x)+\frac{\partial A_3^2(x)}{\partial x} \right )
-\frac{1}{6}\sqrt 3\,\left (A_8^1(x)+\frac{\partial A_8^2(x)}{\partial x} 
\right )
\nonumber\\[2ex]
&& -\frac{1}{2}\,\left (\bar A_3^1(x)+\frac{\partial \bar A_3^2(x)}
{\partial x} \right ) -\frac{1}{2}\sqrt 3\,\left (\bar A_8^1(x)+\frac{\partial 
\bar A_8^2(x)}{\partial x} \right )\Big ] 
\nonumber\\[2ex]
&& +\sin \theta_c \,\cos \theta_c\,\Big [ \left (A_6^1(x)+\frac{\partial 
A_6^2(x)} {\partial x} \right )+\left (\bar A_6^1(x)+\frac{\partial 
\bar A_6^2(x)} {\partial x} \right )\Big ]\,.
\end{eqnarray}
The structure function $g_1^{\rm W^- N}$ is obtained from $g_1^{\rm W^+ N}$
by $\bar A_a^i \rightarrow - \bar A_a^i$ ($i=1,2$).
\begin{eqnarray}
\label{eq:3.19}
g_4^{\rm W^+ N}(x)/x&=&\cos^2 \theta_c \Big [-\frac{4}{3}\, \left (\bar A_0^1(x)
+\frac{\partial \bar A_0^2(x)}{\partial x} \right )
-\frac{2}{3}\sqrt 3\,\left (\bar A_8^1(x)+\frac{\partial \bar A_8^2(x)}
{\partial x} \right )
\nonumber\\[2ex]
&& +2\,\left (A_3^1(x)+\frac{\partial A_3^2(x)}{\partial x}
\right ) \Big ] + \sin^2 \theta_c \Big [- \frac{4}{3}\,\left (\bar A_0^1(x)
+\frac{\partial \bar A_0^2(x)}{\partial x} \right )
\nonumber\\[2ex]
&& - \left ( \bar A_3^1(x)+\frac{\partial \bar A_3^2(x)}{\partial x} \right )
+\frac{1}{3}\sqrt 3\,\left (\bar A_8^1(x)+\frac{\partial \bar A_8^2(x)}
{\partial x} \right )
\nonumber\\[2ex]
&& +\left (A_3^1(x)+\frac{\partial A_3^2(x)} {\partial x} \right ) +\sqrt 3\,
\left (A_8^1(x)+\frac{\partial A_8^2(x)}{\partial x} \right )\Big ]
\nonumber\\[2ex]
&& -2\,\sin \theta_c \,\cos \theta_c\,\Big [ \left (\bar A_6^1(x)+
\frac{\partial \bar A_6^2(x)} {\partial x} \right )+\left (A_6^1(x)
+\frac{\partial A_6^2(x)} {\partial x} \right )\Big ]\,.
\end{eqnarray}
The structure function $g_4^{\rm W^- N}$ is obtained from $g_4^{\rm W^+ N}$
by $A_a^i \rightarrow - A_a^i$ ($i=1,2$).\\
From the definitions of the matrix elements in Eq. (\ref{eq:3.11}) one 
can only obtain results for the following integrals
\begin{eqnarray}
\label{eq:3.20}
&& \int_{-1}^1d\,x V_c^1(x)=V_c^1(0,0)=\Gamma_c \,,\quad 
\int_{-1}^1d\,x A_c^1(x)=A_c^1(0,0)=\Gamma_c^5\,,
\nonumber\\[2ex]
&& \int_{-1}^1d\,x \bar V_c^1(x)=\bar V_c^1(0,0)=0 \,,\quad 
\int_{-1}^1d\,x \bar A_c^1(x)=\bar A_c^1(0,0)=0\,.
\end{eqnarray}
In order to compute the sum rules which are of the type 
$\int_0^1 dx~\Delta F^N(x,q^2)$ one has to convert integrals of the form
$\int_{-1}^1 dx~J_c^k(x)$ into $\int_0^1 dx~J_c^k(x)$. This is only
possible for the following integrals
\begin{eqnarray}
\label{eq:3.21}
&& \int_{-1}^1 dx\, V_c^1(x)=\frac{1}{2}\int_0^1 dx\, V_c^1(x) \,,
\nonumber\\[2ex]
&& \int_{-1}^1d\,x\,\left (A_c^1(x)+\frac{\partial A_c^2(x)}{\partial x}\right )
=\frac{1}{2}\int_0^1d\,x\,A_c^1(x) \,,
\end{eqnarray}
where we have used the property
\begin{eqnarray}
\label{eq:3.22}
\int_{-1}^1d\,x \frac{\partial J_c^k(x)}{\partial x}=0 \,,\qquad J=V,\bar V,A,
\bar A\,.
\end{eqnarray}
However because of the symmetry properties in Eq. (\ref{eq:3.13}) we obtain
\begin{eqnarray}
\label{eq:3.23}
&&\int_{-1}^1 dx\,\bar V_c^1(x)=\int_0^1 dx\,\bar V_c^1(x)+\int_{-1}^0 dx\,
\bar V_c^1(x)=\int_0^1 dx\,\bar V_c^1(x)+\int_0^1 dx\,\bar V_c^1(-x)
\nonumber\\[2ex]
&& =\int_0^1 dx\,\bar V_c^1(x)-\int_0^1 dx\,\bar V_c^1(x)=0\,,
\end{eqnarray}
so that one cannot express $\int_0^1 dx\,\bar V_c^1(x)$ into
$\int_{-1}^1 dx\,\bar V_c^1(x)$. The same holds for 
$\int_0^1d\,x\,\bar A_c^1(x)$.
Therefore we can only compute those sum rules
when the combination of structure functions can be either
expressed into $V_c^1(x)$ (unpolarized scattering) or into
$A_c^1(x)+\frac{\partial A_c^2(x)}{\partial x}$ (polarized scattering).\\ 
Hence for charged current interactions only the following fundamental
sum rules can be derived. They are given by
\begin{eqnarray}
\label{eq:3.24}
&& \mbox{unpolarized Bjorken sum rule \cite{bjork2}}
\nonumber\\[2ex]
&& \int_0^1 d~x\left (F_1^{\rm W^-N}(x,q^2) - F_1^{\rm W^+N}(x,q^2) \right )
\nonumber\\[2ex]
&& \hspace*{4cm}= (I_3^N+\frac{3}{2}\,Y^N) +\cos^2 \theta_c\,(I_3^N
-\frac{3}{2}\,Y^N)\,,
\\[2ex]
\label{eq:3.25}
&& \mbox{Gross Llewellyn Smith sum rule \cite{grll}}
\nonumber\\[2ex]
&& \int_0^1 d~x\left (F_3^{\rm W^-N}(x,q^2) + F_3^{\rm W^+N}(x,q^2) \right )
\nonumber\\[2ex]
&& \hspace*{3cm}= (2\,I_3^N-Y^N+4\,B^N) +\cos^2 \theta_c\,(-2\,I_3^N+3\,Y^N)\,.
\end{eqnarray}
Further we have the polarized analogue of the Bjorken sum rule in 
Eq. (\ref{eq:3.24})
\begin{eqnarray}
\label{eq:3.26}
&& \int_0^1  d~x \left (g_5^{\rm W^-N}(x,q^2) - g_5^{\rm W^+N}(x,q^2) \right )
\nonumber\\[2ex]
&& =\Big  \{-I_3^N(F+D)-\frac{1}{2}\,(3\,F-D)\Big \} +\cos^2 \theta_c\,\Big 
\{-I_3^N(F+D) + \frac{1}{2}(3\,F-D)\Big \}\,.
\nonumber\\
\end{eqnarray}
Because of the Callan-Gross relation mentioned below Eq. (\ref{eq:3.11}) 
the Adler sum rule in Eq. (\ref{eq:2.24}) and its polarized analogue in
Eq. (\ref{eq:2.26}) follow automatically from Eq. (\ref{eq:3.24}) and
Eq. (\ref{eq:3.26}) respectively. Another important feature is that the
Gross-Llewellyn Smith sum rule is determined by the symmetric structure
constants $d_{abc}$ whereas the other charged current sum rules are determined
by the structure constants $f_{abc}$ of the Lie-algebra of $SU(3)_F$.\\
For the flavour group $SU(4)_F$ (see Appendix A) we obtain four additional sum
rules which are not present in the case of $SU(3)_F$. They are given by
\begin{eqnarray}
\label{eq:3.27}
\int_0^1 \frac{d~x}{x}\left (F_2^{\rm W^{\pm} p}(x,q^2) - F_2^{\rm W^{\pm} n}
(x,q^2) \right )=\mp 2\,,
\end{eqnarray}
and
\footnote{Very often Eq. (\ref{eq:3.27}) is called the Adler sum rule.
However this is wrong. The correct one is given in Eq. (\ref{eq:2.24})}
\begin{eqnarray}
\label{eq:3.28}
\int_0^1  \frac{d~x}{x} \left (g_4^{\rm W^{\pm} p}(x,q^2)
- g_4^{\rm W^{\pm} n}(x,q^2) \right ) =\pm 2\,(F'+D')\,.
\end{eqnarray}
The other two sum rules follow from the Callan-Gross relation \cite{cagr}
and they are obtained from the two above via the replacements
$F_2 \rightarrow 2xF_1$ and $g_4 \rightarrow 2xg_5$.
Notice that in the derivation of the expressions above
we have used isospin symmetry which
implies $V_{3,p}^k(x)=-V_{3,n}^k(x)$ and  $V_{c,p}^k(x)=V_{c,n}^k(x)$
for $c\not =3$. In the case of $SU(3)_F$, the expressions in Eqs. 
(\ref{eq:3.27}), (\ref{eq:3.28}) cannot be derived because the combination
of structure functions still contain quantities of the type
$\bar V_3^1(x)$, $\bar A_3^k(x)$ ($k=1,2$). Finally note that $F'$ and $D'$
in Eq. (\ref{eq:3.28}) differ from $F$ and $D$ in Eqs (\ref{eq:2.26}),
(\ref{eq:3.26}) since the latter are only measured when we assume
a $SU(3)_F$ symmetry.

With the help of the light cone algebra we can also derive sum rules
for the structure functions measured in neutral current processes.
If we define $s=\sin \theta_W$, $c=\cos \theta_W$, where $\theta_W$ denotes
the weak angle, the neutral electroweak currents for $SU(3)_F$ are given by
\begin{eqnarray}
\label{eq:3.29}
&& J_{\gamma}^{\mu}(y)= V_{\gamma}^{\mu}(y)\,,
\nonumber\\[2ex]
&& J_Z^{\mu}(y)=(1-2\,s^2)\,V_{\gamma}^{\mu}(y) -A_{\gamma}^{\mu}(y) 
-\frac{1}{3}\,\Big (V_0^{\mu}(y)-A_0^{\mu}(y)\Big )\,,
\nonumber\\[2ex]
\mbox{with}&& V_{\gamma}^{\mu}(y) \equiv V_3^{\mu}(y)
+\frac{1}{3}\sqrt 3\,V_8^{\mu}(y) \quad
A_{\gamma}^{\mu}(y) \equiv A_3^{\mu}(y)+\frac{1}{3}\sqrt 3\,A_8^{\mu}(y)\,.
\end{eqnarray}
Substitution of these currents into the hadronic structure tensor in 
Eq. (\ref{eq:2.2}) yields the following results
\begin{eqnarray}
\label{eq:3.30}
W_{\gamma\gamma}^{\mu\nu}&=&W_{33,VV}^{\mu\nu}+\frac{2}{3}\sqrt 3\,
W_{\{38\}} ^{\mu\nu}+\frac{1}{3}\,W_{88,VV}^{\mu\nu}\,,
\\[2ex]
\label{eq:3.31}
W_{ZZ}^{\mu\nu}&=&2(1-2\,s^2+2\,s^4)\Big [W_{33,VV}^{\mu\nu}+\frac{2}{3}
\sqrt 3\, W_{\{38\},VV}^{\mu\nu}+\frac{1}{3}\,W_{88,VV}^{\mu\nu}\Big ]
\nonumber\\[2ex]
&& -\frac{4}{3}(1-s^2)\Big [ W_{\{30\},VV}^{\mu\nu}+\frac{1}{3}\sqrt 3\,
W_{\{80\},VV}^{\mu\nu}\Big ]+\frac{2}{9}\,W_{00,VV}^{\mu\nu}
\nonumber\\[2ex]
&& -2(1-2\,s^2)\Big [W_{33,AV}^{\mu\nu}+\frac{2}{3}\sqrt 3\,
W_{\{38\},AV}^{\mu\nu} +\frac{1}{3}\,W_{88,AV}^{\mu\nu}\Big ]
\nonumber\\[2ex]
&& +\frac{4}{3}(1-s^2)\Big [ W_{\{30\},AV}^{\mu\nu} +\frac{1}{3}\sqrt 3\,
W_{\{80\},AV}^{\mu\nu} \Big ] -\frac{2}{9}\,W_{00,AV}^{\mu\nu}\,,
\\[2ex]
\label{eq:3.32}
W_{\gamma Z+Z \gamma}^{\mu\nu}&=&2(1-2\,s^2)\Big [W_{33,VV}^{\mu\nu}
+\frac{2}{3}\sqrt 3\, W_{\{38\},VV}^{\mu\nu}+\frac{1}{3}\,
W_{88,VV}^{\mu\nu}\Big ] -\frac{2}{3}\Big [ W_{\{30\},VV}^{\mu\nu}
\nonumber\\[2ex]
&& +\frac{1}{3}\sqrt 3\, W_{\{80\},VV}^{\mu\nu}\Big ]-2\Big [
W_{33,AV}^{\mu\nu}+\frac{2}{3}\sqrt 3\, W_{\{38\},AV}^{\mu\nu}+\frac{1}{3}\,
W_{88,AV}^{\mu\nu}\Big ]
\nonumber\\[2ex]
&& +\frac{2}{3}\Big [ W_{\{30\},AV}^{\mu\nu}+\frac{1}{3}
\sqrt 3\, W_{\{80\},AV}^{\mu\nu} \Big ]\,.
\end{eqnarray}
Since the above structure tensor only involve anti-commutators, indicated
by $\{\}$, the structure functions are determined by the values for $d_{abc}$.
Let us introduce the following shorthand notations defined by
\begin{eqnarray}
\label{eq:3.33}
V_{\gamma}(x)&=&V_3^1(x)+\frac{1}{3}\sqrt 3V_8^1(x)\,,
\nonumber\\[2ex]
A_{\gamma}(x)&=&A_3^1(x)+\frac{\partial A_3^2(x)}{\partial x}
+\frac{1}{3}\sqrt 3\left (A_8^1(x)+\frac{\partial A_8^2(x)} {\partial x}\right )
\,,
\nonumber\\[2ex]
V_{\gamma \gamma}(x)&=&\frac{4}{9}\,V_0^1(x)+\frac{1}{3}\,V_3^1(x) 
+\frac{1}{9}\sqrt 3\,V_8^1(x)\,,
\nonumber\\[2ex]
A_{\gamma \gamma}(x)&=&\frac{4}{9}\left (A_0^1(x)+\frac{\partial A_0^2(x)}
{\partial x}\right )+\frac{1}{3}\left (A_3^1(x)+\frac{\partial A_3^2(x)}
{\partial x}\right ) 
\nonumber\\[2ex]
&& +\frac{1}{9}\sqrt 3\left (A_8^1(x)+\frac{\partial A_8^2(x)} {\partial x}
\right )\,.
\end{eqnarray}
The neutral current structure functions read as follows
\begin{eqnarray}
\label{eq:3.34}
F_2^{\rm \gamma N}(x)&=& x\,\bar V_{\gamma \gamma}(x)\,,
\\[2ex]
\label{eq:3.35}
g_1^{\rm \gamma N}(x)&=& \frac{1}{2}\,A_{\gamma \gamma}(x)\,,
\\[2ex]
\label{eq:3.36}
F_2^{\rm Z N}(x)&=& 2\,x\,(1-2\,s^2+2\,s^4)\,\bar V_{\gamma \gamma}(x)
-\frac{2}{3}\,x\,(1-s^2)\,\bar V_{\gamma}(x) +\frac{1}{9}\,x\,\bar V_0^1(x)\,,
\\[2ex]
\label{eq:3.37}
F_3^{\rm Z N}(x)&=& 2\,(1-2\,s^2)\,V_{\gamma \gamma}(x)
-\frac{2}{3}(1-s^2)\,V_{\gamma}(x) +\frac{1}{9}\,V_0^1(x)\,,
\\[2ex]
\label{eq:3.38}
g_1^{\rm Z N}(x)&=& (1-2\,s^2+2\,s^4)\,A_{\gamma \gamma}(x)
-\frac{1}{3}(1-s^2)\,A_{\gamma}(x) 
\nonumber\\[2ex]
&& +\frac{1}{18}\,\left(A_0^1(x) +\frac{\partial A_0^2(x)} {\partial x}\right )
\,,
\\[2ex]
\label{eq:3.39}
g_4^{\rm Z N}(x)&=& -2\,x\,(1-2\,s^2)\,\bar A_{\gamma \gamma}(x)
+\frac{2}{3}\,x\,(1-s^2)\,\bar A_{\gamma}(x) 
\nonumber\\[2ex]
&& -\frac{1}{9}\,x\,\left(\bar A_0^1(x)
+\frac{\partial \bar A_0^2(x)} {\partial x}\right )\,,
\\[2ex]
\label{eq:3.40}
F_2^{\rm \gamma Z,N}(x)&=& 2\,x\,(1-2\,s^2)\,\bar V_{\gamma \gamma}(x)
-\frac{1}{3}\,x\,\bar V_{\gamma}(x) \,,
\\[2ex]
\label{eq:3.41}
F_3^{\rm \gamma Z,N}(x)&=& 2\,V_{\gamma \gamma}(x) - \frac{1}{3}\,V_{\gamma}(x)
\,,
\\[2ex]
\label{eq:3.42}
g_1^{\rm \gamma Z,N}(x)&=& (1-2\,s^2)\,A_{\gamma \gamma}(x)
-\frac{1}{6}\,A_{\gamma}(x)\,,
\\[2ex]
\label{eq:3.43}
g_4^{\rm \gamma Z,N}(x)&=& -2\,x\,\bar A_{\gamma \gamma}(x)
+\frac{1}{3}\,x\, \bar A_{\gamma}(x)\,.
\end{eqnarray}
If the proton is replaced by the neutron, the structure functions of the latter
are derived from the former via the substitution $J_3^i \rightarrow -J_3^i$ 
with $i=1,2$ and $J=V, \bar V, A, \bar A$. As has been mentioned below
Eq. (\ref{eq:3.23}), we can derive sum rules when the structure functions
contain the matrix elements of $V_c^1(x)$ and $A_c^1(x)
+\frac{\partial A_c^2(x)}{\partial x}$ only. The results are given by
\begin{eqnarray}
\label{eq:3.44}
\int_0^1 dx\,F_3^{\rm Z N}(x,q^2) &=& \frac{3}{2}\,(1-2\,s^2)\,B^N
-\frac{1}{3}\,s^2\,(2\,I_3^N+S^N) \,,
\\[2ex]
\label{eq:3.45}
\int_0^1 dx\,F_3^{\rm \gamma Z,N}(x,q^2)&=&\frac{3}{2}\,B^N
+\frac{1}{6}\,(2\,I_3^N+S^N)\,.
\end{eqnarray}
For the longitudinal spin structure function $g_1$ we obtain
\begin{eqnarray}
\label{eq:3.46}
\int_0^1 dx\,g_1^{\rm \gamma N}(x,q^2)&=&f_{\gamma}^{\rm S}+f^{\rm NS}\,,
\\[2ex]
\label{eq:3.47}
\int_0^1 dx\,g_1^{\rm Z N}(x,q^2) &=&f_{\rm Z}^{\rm S}-2\,s^2(1-2\,s^2)\,
f^{\rm NS} \,,
\\[2ex]
\label{eq:3.48}
\int_0^1 dx\,g_1^{\rm \gamma Z,N}(x,q^2) &=& f_{\rm \gamma Z}^{\rm S}\,
+ (1-4\,s^2)\,f^{\rm NS} \,,
\end{eqnarray}
\begin{eqnarray}
\label{eq:3.49}
f^{\rm NS}&=&\frac{1}{6}\,I_3^N\,(F+D)+\frac{1}{36}\,(3\,F-D)\,,
\\[2ex]
\label{eq:3.50}
f_{\gamma}^{\rm S}&=&\frac{2}{9}\,\Gamma_0^{5,N}\,,
\nonumber\\[2ex]
f_{\rm Z}^{\rm S}&=&(\frac{4}{9}-\frac{8}{9}\,s^2+\frac{8}{9}s^4)\,
\Gamma_0^{5,N} \,,
\nonumber\\[2ex]
f_{\gamma Z}^{\rm S}&=&\frac{4}{9}(1-2\,s^2)\,\Gamma_0^{5,N} \,.
\end{eqnarray}
Here the superscripts $S$ and $NS$ refer to the singlet and non-singlet
representations of the flavour group $SU(3)_F$. Contrary to 
structure function $F_3$ in Eqs. (\ref{eq:3.44}), (\ref{eq:3.45}),
the sum rules for $g_1$  Eqs. (\ref{eq:3.46})-(\ref{eq:3.48}) cannot be
expressed into the quantum numbers of the hadron because $\Gamma_0^{5,N}$
in Eq. (\ref{eq:3.50}) is unknown. Expressions (\ref{eq:3.46})-(\ref{eq:3.48})
are generalizations of the Ellis-Jaffe sum rule for electro-weak currents
which was originally derived for electro-production \cite{elja}.
Moreover the singlet axial-vector current $A_0^{\mu}(y)$ leading to the matrix
element $A_0^1(x)+\frac{\partial A_0^2(x)}{\partial x}$ in Eq. (\ref{eq:3.33})
is not conserved due to the Adler-Bell-Jackiw (ABJ)anomaly \cite{abj}. 
Therefore
$\Gamma_0^{5,N}$ will become scale dependent when higher order QCD corrections
are included. Hence the Ellis-Jaffe sum rule is non-fundamental and it is
no surprise that it is violated experimentally \cite{ash}. However using 
isospin symmetry one can derive the generalization of the polarized
Bjorken sum rule \cite{bjork1} given by the expressions
\begin{eqnarray}
\label{eq:3.51}
\int_0^1 dx \left (g_1^{\gamma p}(x,q^2) - g_1^{\gamma n}
(x,q^2) \right )=\frac{1}{6}\,(F+D)\,,
\\[2ex]
\label{eq:3.52}
\int_0^1 dx\left (g_1^{Z p}(x,q^2) - g_1^{Z n}
(x,q^2) \right )=-\frac{1}{3}\,s^2(1-2\,s^2)\,(F+D)\,,
\\[2ex]
\label{eq:3.53}
\int_0^1 dx\left (g_1^{\gamma Z,p}(x,q^2) - g_1^{\gamma Z,n}
(x,q^2) \right )=\frac{1}{6}\,(1-4\,s^2)\,(F+D)\,.
\end{eqnarray}
When in the Bjorken limit the leading contributions are coming from twist two 
operators only, the structure functions can be expressed into parton densities
which become scale dependent when higher order QCD corrections are included.
Hence one obtains relations between all bilocal operator matrix elements and
the parton densities so that one can also compute sum rules involving the
matrix elements $\bar V_c^k(x)$ or $\bar A_c^k(x)$. However the integrals
$\int_0^1dx~\bar V_c^k(x)$, $\int_0^1dx~\bar A_c^k(x)$ can only be 
computed if one makes additional model dependent assumptions. Moreover it
turns out they receive scaling violating contributions as e.g. is observed
for the Ellis-Jaffe sum rule (see below Eq. (\ref{eq:3.50})). Therefore it
is no surprise that the results are very often in disagreement with experiment
as we will show in an example below. Therefore these sum rules will be called
non-fundamental or parton model sum rules.
The relations between the bilocal operator matrix elements and the 
unpolarized parton densities $q(x)$ are given by
\begin{eqnarray}
\label{eq:3.54}
V_3^1(x)&=&\frac{1}{2}\Big (V_u(x)-V_d(x)\Big )\,,
\nonumber\\[2ex]
V_8^1(x)&=&\frac{1}{2\sqrt 3}\Big (V_u(x)+V_d(x)-2\,V_s(x)\Big )\,,
\nonumber\\[2ex]
V_0^1(x)&=&\frac{1}{2}\Big (V_u(x)+V_d(x)+V_s(x)\Big )\,,
\nonumber\\[2ex]
\bar V_3^1(x)&=&\frac{1}{2}\Big (\Delta_u(x)-\Delta_d(x)\Big)\,,
\nonumber\\[2ex]
\bar V_8^1(x)&=&\frac{1}{2\sqrt 3}\Big (\Delta_u(x)+\Delta_d(x)-2\,\Delta_s(x)
\Big )\,,
\nonumber\\[2ex]
\bar V_0^1(x)&=&\frac{1}{2} \Sigma(x)\,.
\end{eqnarray}
Here the non-singlet $V_q(x),~\Delta_q(x)$ and the singlet $\Sigma(x)$ parton
densities are defined by
\begin{eqnarray}
\label{eq:3.55}
&& V_q(x)=q(x)-\bar q(x)\,, \qquad \Delta_q(x)=q(x)+\bar q(x)-\frac{1}{3}
\Sigma(x) \,,
\nonumber\\[2ex]
&&\Sigma(x)=\sum_{q=u,d,s} q(x)+\bar q(x)\,.
\end{eqnarray}
For the longitudinally polarized parton densities $\delta q(x)$ we obtain 
\begin{eqnarray}
\label{eq:3.56}
A_3^1(x)+\frac{\partial A_3^2(x)}{\partial x}&=&\frac{1}{2}\Big (
\Delta_{\delta u}(x) - \Delta_{\delta d}(x)\Big )\,,
\nonumber\\[2ex]
A_8^1(x)+\frac{\partial A_8^2(x)}{\partial x}&=&\frac{1}{2\sqrt 3}\Big (
\Delta_{\delta u}(x)+\Delta_{\delta d}(x)-2\,\Delta_{\delta s}(x)\Big )\,,
\nonumber\\[2ex]
A_0^1(x)+\frac{\partial A_0^2(x)}{\partial x}&=&\frac{1}{2}\delta \Sigma (x)
\,,
\nonumber\\[2ex]
\bar A_3^1(x)+\frac{\partial \bar A_3^2(x)}{\partial x}&=&\frac{1}{2}
\Big (V_{\delta u}(x)-V_{\delta d}(x)\Big )\,,
\nonumber\\[2ex]
\bar A_8^1(x)+\frac{\partial \bar A_8^2(x)}{\partial x}&=&\frac{1}{2\sqrt 3}
\Big (V_{\delta u}(x)+V_{\delta d}(x) -2\,V_{\delta s}(x)\Big )\,,
\nonumber\\[2ex]
\bar A_0^1(x)+\frac{\partial \bar A_0^2(x)}{\partial x}&=&\frac{1}{2}\Big (
V_{\delta u}(x)+V_{\delta d}(x) +V_{\delta s}(x)\Big ) \,,
\end{eqnarray}
with similar notations as in Eq. (\ref{eq:3.55}) for the polarized quark 
densities. If one includes higher order QCD corrections all parton
densities will depend on a scale except if one takes the first moment
represented by the integrals of the type $\int_0^1 dx~V_c^1(x)$ or
$\int_0^1 dx~A_c^1(x)$. This is because the integrals are related to
conserved vector and axial-vector currents respectively. Hence from 
Eqs. (\ref{eq:3.54}), (\ref{eq:3.56}) it follows that
$\int_0^1dx~V_q(x)$ and $\int_0^1dx~\Delta_{\delta q}(x)$ are scale independent
This does not hold for $\int_0^1dx~\Delta_q(x)$ and 
$\int_0^1dx~V_{\delta q}(x)$. Here the scale dependence
is ruled by an anomalous dimension which becomes non-vanishing
in order $\alpha_s^2$ (see \cite{rosa}). This for instance happens in the 
Gottfried sum rule \cite{gott} given by
\begin{eqnarray}
\label{eq:3.57}
\int_0^1 \frac{d~x}{x}\left (F_2^{\rm \gamma p}(x,q^2) - F_2^{\gamma n}
(x,q^2) \right )=\frac{2}{3}\int_0^1dx\,\bar V_3^1(x)\,,
\end{eqnarray}
where $\bar V_3^1(x)=(\Delta_u(x)-\Delta_d(x))/2$ (see Eq. (\ref{eq:3.54})).
Hence this expression acquires second order QCD corrections containing scaling
violating terms. Furthermore the right-hand side is model dependent 
because it is not related to the expectation value of a conserved (axial-) 
vector current.
Only under certain assumptions i.e. $\bar u(x)=\bar d(x)$ the above integral 
yields $1/3$. Because of the scaling violating term and the model dependence
it is no surprise that this result is in disagreement with experiment 
(see e.g \cite{amadr}). 
Hence one can conclude that all sum rules which cannot be related to the
expectation values of (axial-) vector currents are model dependent and
show scaling violating terms in the perturbation series due to non-vanishing 
anomalous dimensions. These sum rules are non-fundamental and they are not 
suitable as a test of perturbative QCD. A collection of these parton model
sum rules is shown in table 2 of \cite{blko2}.

There is a second class of non-fundamental sum rules which originate
from structure functions which contain besides twist two also twist
three contributions. They only appear in polarized scattering and
hold for charged current as well as neutral current processes. 
These sum rules can be derived from Eq. (\ref{eq:3.10}) and Eq. (\ref{eq:3.22}).
The most well known one is the Burkhardt-Cottingham sum rule given by
\cite{buco}
\begin{eqnarray} 
\label{eq:3.58}
\int_0^1dx\,g_{2,ab}^{VV}(x,q^2)=\int_0^1dx\,g_{2,ab}^{AA}(x,q^2)=0\,.
\end{eqnarray}
However the sum rule above is
not the only one which equals zero. In a similar way one can also derive that
\begin{eqnarray}
\label{eq:3.59}
\int_0^1dx\,\left (g_{4,ab}^{AV}(x,q^2)-g_{3,ab}^{AV}(x,q^2)\right )=0\,,
\end{eqnarray}
\begin{eqnarray}
\label{eq:3.60}
\int_0^1dx\,\left (2\,x\,g_{5,ab}^{AV}(x,q^2)-g_{3,ab}^{AV}(x,q^2)\right )=0\,,
\end{eqnarray}
yield zero. For charged current interactions where only the expectation value
of the axial vector current shows up one obtains
\begin{eqnarray}
\label{eq:3.61}
\int_0^1\frac{dx}{x}\,\left (g_{3,[ab]}^{AV}(x,q^2)
+2\,x\,g_{9,[ab]}^{AV}(x,q^2)\right )= \,f_{abc}\,\Gamma_c^5\,,
\end{eqnarray}
\begin{eqnarray}
\label{eq:3.62}
\int_0^1\frac{dx}{x}\,\left (g_{3,[ab]}^{AV}(x,q^2)
-2\,x\,g_{8,[ab]}^{AV}(x,q^2)\right )= \,f_{abc}\,\Gamma_c^5\,.
\end{eqnarray}
They look like the expressions derived in Eq. (\ref{eq:2.14}) and in
the next section we will study whether they receive QCD and power corrections.
Notice that the last two sum rules are hard to measure since they contain the
structure functions $g_8,g_9$ which do not show up in the cross section
when lepton masses are neglected.

\mysection{QCD and power corrections to the sum rules}
In this section we will discuss how the sum rules derived in the previous 
section are modified by QCD and power corrections. We will show that when
power corrections occur the sum rule also receives QCD corrections.
This statement only holds for twist two sum rules. At the end we also
discuss sum rules which receive twist three contributions. All sum rules
derived in the previous sections can be written as
\begin{eqnarray}
\label{eq:4.1}
\int_0^1dx\,\Delta F_i(x,q^2,m^2)=\sum_a \Gamma_a^N\,
{\cal C}_{i,q}\left (\alpha_s(\mu^2),\frac{q^2}{m^2}\right )\quad + \quad 
\mbox{higher twist}\,.
\end{eqnarray}
The above form follows from the operator expansion where
$\Gamma_a^N$ is the expectation value of a vector current sandwiched between 
the hadronic state $N(p,s)$ (see Eq. (\ref{eq:2.10})). In the case of an
axial-vector current $\Gamma_a^N$ is replaced by $\Gamma_a^{5,N}$. 
The quantity ${\cal C}_{i,q}$ denotes the first moment of the quark
non-singlet coefficient function and it contains all QCD and 
power corrections of the type $m^2/q^2$. Notice that vector currents are
conserved so that they are not renormalized. However axial-vector currents are 
partially conserved which means that they receive finite renormalizations. 
Examples of this phenomenon are the isospin currents $V_3^{\mu}$
and $A_3^{\mu}$. The expectation values read
\begin{eqnarray}
\label{eq:4.2}
\Gamma_3^N=2\,I_3^N\,g_V  \,,\qquad \Gamma_3^{5,N}=2\,I_3^N\,g_A\,.
\end{eqnarray}
Here we have $g_V=1$ but $g_A=F+D\not =1$ (see below Eq. (\ref{eq:2.27})).
In both cases $\Gamma_a^N$ as well as $\Gamma_a^{5,N}$ are scale independent
so that all power corrections can be either attributed to ${\cal C}_q$ or
to higher twist. When the singlet axial-vector current shows up, 
as happens for the Ellis-Jaffe sum rule, the corresponding expectation
value $\Gamma_0^{5,N}$ becomes scale dependent and the coefficient function
receives logarithmic corrections \cite{zn1}. This is because the symmetry
is so badly broken by the ABJ-anomaly \cite{abj} that the quantity
$A_0^{\mu}$ will receive infinite renormalizations. 

For unpolarized scattering the first moment of the coefficient function,
presented on the right-hand side of Eq. (\ref{eq:4.1}), is given by
\begin{eqnarray}
\label{eq:4.3}
\int_0^1dz\,\hat {\cal F}_i(z,q^2,m^2)=\Gamma_q\,{\cal C}_{i,q}(\alpha_s,
\frac{q^2}{m^2})\,.
\end{eqnarray}
In the case of polarized scattering we have
\begin{eqnarray}
\label{eq:4.4}
\int_0^1dz\,\hat g_i(z,q^2,m^2)=\Gamma_q^5\,\delta {\cal C}_{i,q}(\alpha_s,
\frac{q^2}{m^2})\,.
\end{eqnarray}
Here $\hat {\cal F}_i$ and $\hat g_i$ are the partonic structure functions
which depend on the mass $m$, virtuality $q^2$ of the vector-boson and the 
partonic scaling variable $z=-q^2/2p\cdot q$ where $p$ now stands for the 
momentum
of the incoming quark. The partonic quantities are defined in the same way 
as the hadronic structure functions in Eq. (\ref{eq:2.3}) except that the 
hadronic currents are replaced by the quark currents. 
Further we have the properties
\begin{eqnarray}
\label{eq:4.5}
\Gamma_q =1 \,,\qquad \Gamma_q^5=\sum_{n=0}^{\infty} \Gamma_q^{5,n}\,
\left (\frac{\alpha_s(\mu^2)}{4\,\pi} \right)^n \,,\qquad \Gamma_q^{5,0} =1\,.
\end{eqnarray}
The results above follow from quark current conservation and partial
conservation of the quark axial-vector current ($\partial_{\mu}A_a^{\mu}
\not=0$) which happens if the quark has a mass $m\not=0$. In the case the quark
is massless the axial vector is conserved too and we have
$\Gamma_q^5=1$ in all orders of perturbation theory.\\ 
The partonic structure functions are computed from the process
\begin{eqnarray}
\label{eq:4.6}
V + q \rightarrow 'partons' \,,\quad \mbox{with} \quad V=\gamma,Z,W^{\pm}\,,
\end{eqnarray}
where $'partons'$ represents all (anti-) quarks and gluons which can be
produced in the final state. 
Further the general vertex for the coupling of the vector boson $V$
to the quarks will be denoted by
\begin{eqnarray}
\label{eq:4.7}
\Gamma_{a,\mu}^{V,(0)}&=&-i~({\it v}_a^V + {\it a}_a^V \gamma_5) \gamma_\mu
\,, \qquad V=\gamma, Z, W\,,
\end{eqnarray}
where
\begin{eqnarray}
\label{eq:4.8}
{\it v}_a^{\gamma}&=& Q_a \,,\qquad {\it a}_a^{\gamma}= 0\,,
\nonumber\\[2ex]
{\it v}_a^Z & =& T^3_a -2\,s_w^2 Q_a \,,\qquad {\it a}_a^Z  = T^3_a\,,
\nonumber\\[2ex]
&& {\it v}_a^W={\it a}_a^W=1\,.
\end{eqnarray}
In the case of the quarks the electroweak charges are equal to
\begin{eqnarray}
\label{eq:4.9}
Q_a&=&\frac{2}{3}, \quad T_3^a=\frac{1}{2}\,, \quad a=u,c,t\,,
\nonumber\\[2ex]
Q_a&=&-\frac{1}{3}, \quad T_3^a=-\frac{1}{2}\,, \quad a=d,s,b\,.
\end{eqnarray}
In this section we aim to compute the coefficient
function up to order $\alpha_s$ for non-zero masses of the quarks.
We only show explicit results for those structure functions
which show up in the cross section for vanishing lepton masses which
means that no formulae are given for $\hat {\cal F}_4,\hat {\cal F}_5$, 
$\hat g_6,\hat g_8,\hat g_9$. In lowest order we have the process
\begin{eqnarray}
\label{eq:4.10}
V(q) ~+~ q(p,s)\rightarrow q(p')\,, \quad \mbox{with} \quad p^2=p'^2=m^2\,,
\end{eqnarray}  
which provides us with the Born approximations to the unpolarized and
polarized coefficient functions denoted by ${\cal C}_{i,q}^{(0)}$ and 
$\delta{\cal C}_{i,q}^{(0)}$ respectively. 
In this paper the calculations are only performed for neutral current 
interactions for which $p^2= p'^2$ 
\footnote{Sum rules for $p^2=0$ and $p'^2=m^2$ are treated up to first
order in \cite{blne}}. However the results are such that the conclusions
also hold for other mass assignments where $p^2\not = p'^2$ which occurs
for charged current processes. There is only one exception. In the case
of the mass assignment in Eq. (\ref{eq:4.10}) it turns out that
$g_6^{VV}=0$ whereas $g_6^{AA}\not =0$ which is in disagreement with the
result found for the bilocal current algebra in Eq. (\ref{eq:3.10}).
This is not surprising because the vector current is conserved in
contrast to the axial vector current. We checked that for the choice
$p^2=m^2$ and $p'^2=0$, where both currents are not conserved, one
obtains the relation $g_6^{VV}=g_6^{AA}$ which is in agreement
with Eq. (\ref{eq:3.10}). For all other structure functions 
the mass assignment is irrelevant provided one takes the Bjorken limit.
In next order we receive contributions from gluon bremsstrahlung given by
\begin{eqnarray}
\label{eq:4.11}
V(q) ~+~ q(p,s)\rightarrow q(p') + g(k)\,,
\end{eqnarray}  
and the virtual corrections to Eq. (\ref{eq:4.10}).  
Because of the infrared divergence appearing at $z=1$ we have to split the 
integrals in Eqs. (\ref{eq:4.3}), (\ref{eq:4.4}) into
\begin{eqnarray}
\label{eq:4.12}
\int_0^1dz\,\hat {\cal F}_{i,q}^{(1)}(z,q^2,m^2)&=&\int_0^{z_{max}}dz\,\hat 
{\cal F}_{i,q}^{\rm HARD}(z,q^2,m^2) 
+\int_{z_{max}}^1dz\,\hat {\cal F}_{i,q}(z,q^2,m^2,\lambda^2) 
\nonumber\\[2ex]
&& +\int_0^1dz\,\hat {\cal F}_{i,q}^{\rm VIRT}(z,q^2,m^2,\lambda^2)\,,
\nonumber\\[2ex]
\hat {\cal F}_{i,q}^{\rm HARD}(z,q^2,m^2)&\equiv&
\hat {\cal F}_{i,q}(z,q^2,m^2,0)\,,
\nonumber\\[2ex]
z_{max}&=&\frac{-q^2}{-q^2+2~m~\omega}\,, \quad \mbox{with} \quad \omega \ll m
\,,
\end{eqnarray}
with a similar expression for Eq. (\ref{eq:4.4}). Here we have introduced
a gluon mass $\lambda$ in order to regularize the infrared divergence. The
computation of the hard gluon, soft gluon and virtual gluon part of
the structure functions proceed in the same way as outlined in \cite{hune}
where the calculation was carried out for $-q^2 \gg m^2$. The computation
of the hard gluon part ${\cal F}_i^{\rm HARD}$ is straightforward and the 
results are given in Appendix B. The soft gluon integral is given by
\begin{eqnarray}
\label{eq:4.13}
\int_{z_{max}}^1dz\,\hat {\cal F}_{i,q}(z,q^2,m^2,\lambda^2)
=S^{\rm SOFT}(q^2,m^2,\lambda^2,\omega) \,{\cal C}_{i,q}^{(0)}\,,
\end{eqnarray}
where $S^{\rm SOFT}$ is given by
\begin{eqnarray}
\label{eq:4.14}
S^{SOFT}&=& -{\alpha_s \over 8 \pi^2} C_F
           \int_0^{\omega} {d^3k \over k^0}\left(
           {m^2 \over (p \cdot k)^2}+{m^2 \over (p' \cdot k)^2}
           -{2 p\cdot p' \over (p\cdot k)(p' \cdot k) } \right)\,,
\end{eqnarray}
and $C_F$ denotes the colour factor which in QCD equals to $4/3$.
For $\lambda^2\ll 2\,m\,\omega \ll m^2$ one obtains the result
\begin{eqnarray}
\label{eq:4.15}
S^{SOFT}&=&{\alpha_s \over 4 \pi} C_F \Bigg[
         -2 \ln\left({4 \omega^2 \over \lambda^2}\right)
         +2 -{4 m^2 -2 q^2 \over \sqrq } \Bigg \{
         -\ln \left( {4 \omega^2 \over \lambda^2 }\right) \ln(t)
\nonumber \\ [2ex]
&& -2 {\cal L}i_2 \left({1 \over t}\right)
         -2 {\cal L}i_2 \left(-{1 \over t}\right)
         -3 \ln^2(t)-\ln(t) +2 \ln(t) \ln(t-1)
\nonumber \\ [2ex]
&&          +2 \ln(t) \ln(t+1)+\zeta(2) \Bigg \} \Bigg]\,,
\nonumber\\[2ex]
t &=& { \sqrt{q^4 -4 m^2 q^2} - q^2 \over
       \sqrt{q^4 -4 m^2 q^2} + q^2 }\,.
\end{eqnarray}
Here ${\cal L}i_2(z)$ denotes the dilogarithm which is defined in \cite{lewin}.
The virtual gluon part is obtained from the order $\alpha_s$ corrected 
vector-boson quark vertex given by
\begin{eqnarray}
\label{eq:4.16}
\Gamma_{q,\mu}^{V,(1)}&=&-i\Bigg[ \gamma_\mu (1+ {\cal R}_1)~{\it v}_q^V+
\gamma_5 \gamma_\mu (1+ {\cal R}_1 +2 {\cal R}_2)~{\it a}_q^V 
\nonumber\\[2ex]
&&+{(p_{\mu}+{p'}_\mu)\over 2 m} {\cal R}_2~ {\it v}_q^V+
\gamma_5 {q_\mu\over 2 m} {\cal R}_3~ {\it a}_q^V \Bigg ]\,.
\end{eqnarray}
For $\lambda^2 \ll m^2$ the functions ${\cal R}_i$ become equal to
\begin{eqnarray}
\label{eq:4.17}
{\cal R}_1&=& {\alpha_s \over 4 \pi} C_F \Bigg[
            \left( {4 m^2 -2 q^2 \over \sqrq} \ln(t) -2\right) 
            \ln\left({\lambda^2 \over m^2}\right) -4
            -3 {\sqrq \over q^2} \ln(t)
\nonumber \\ [2ex]
&& +{ 4 m^2 -2 q^2 \over \sqrq} \Bigg \{ {3 \over 2} \ln^2(t) -2 \ln(t) \ln(t+1)
   +2 {\cal L}i_2\left (-{ 1 \over t} \right) + \zeta(2)\Bigg\} \Bigg]\,,
\nonumber\\ [2ex]
{\cal R}_2&=& {\alpha_s \over 4 \pi}  C_F\Bigg[
     -{4 m^2 \over \sqrq} \ln(t) \Bigg]\,,
\nonumber\\ [2ex]
{\cal R}_3&=& {\alpha_s \over 4 \pi}  C_F\Bigg[
     -\left( 3 -{4 m^2 \over q^2} \right) {4 m^2 \over \sqrq} \ln(t) 
     -{8 m^2 \over q^2} \Bigg] \,.
\end{eqnarray}
The above expressions also provide us with the renormalization constants
of the vector and axial vector current since
\begin{eqnarray}
\label{eq:4.18}
i\,\Gamma_{q,\mu}^V(q)={\it v}_q^V\langle q(p',s) \mid V_{\mu} \mid
q(p,s) \rangle -{\it a}_q^V\langle q(p',s) \mid A_{\mu} \mid q(p,s) \rangle\,.
\end{eqnarray}
For $p=p'$ or $q^2=0$ we obtain
\begin{eqnarray}
\label{eq:4.19}
\Gamma_q=1+{\cal R}_1+{\cal R}_2=1 \,,\quad \Gamma_q^5=1+{\cal R}_1+2\,
{\cal R}_2 =1-\frac{\alpha_s(\mu^2)}{4\,\pi}\,C_F\,\Big \{2\Big \}\,.
\end{eqnarray}
The computation of $\hat {\cal F}_{i,q}^{\rm VIRT}$ proceeds in the same way
as for the Born approximations ${\cal F}_{i,q}^{(0)}$. Since they are all
proportional to $\delta(1-z)$ the integral in Eq. (\ref{eq:4.12}) becomes
trivial. After adding this result to the soft gluon integral in 
Eq. (\ref{eq:4.13}) the infrared regulator $\lambda$ will be cancelled.
The resulting expression is given by
\begin{eqnarray}
\label{eq:4.20}
{\cal T}_{i,q}^{\rm S+V}=S^{\rm SOFT}(q^2,m^2,\lambda^2,\omega) 
\,{\cal C}_{i,q}^{(0)}
+\int_0^1dz\,\hat {\cal F}_{i,q}^{\rm VIRT}(z,q^2,m^2,\lambda^2)\,.
\end{eqnarray}
Using the expressions for $\hat {\cal F}_{i,q}^{\rm HARD}$ in Appendix B
one can now compute the first integral on the right-hand side in 
Eq. (\ref{eq:4.12}). After taking the limit $\omega \rightarrow 0$ we
obtain the following sum rules
\begin{eqnarray}
\label{eq:4.21}
{\cal C}_{1,q}
&=& \int_0^1 dz \hat {\cal F}_{1,q}(z,q^2,m^2) = 
\nonumber\\[2ex]
&& {\it v}_q^{V_1}{\it v}_q^{V_2}\Bigg [\frac{1}{2} +\frac{\alpha_s(\mu^2)}
{4\pi} C_F \Bigg \{ \Bigg(- 12 + {4 q^6 \over \mpq^3 } - {6 q^4 \over \mpq^2}
\nonumber \\[2ex]
&&   +{5 q^2 \over \mpq }\Bigg)\lnsqt +{1 \over \sqrq} \Bigg (16 m^2 - 5 q^2 
\Bigg ) \ln(t) 
\nonumber \\[2ex]
&&      -{2 q^4 \over \mpq^2 }  +{2 q^2 \over \mpq }
      +\left ( 6 m^2 -{q^2 \over 2} \right) {\cal J} \Bigg \}  \Bigg ] 
\nonumber \\[2ex]
&& +{\it a}_q^{V_1}{\it a}_q^{V_2} \Bigg [\frac{1}{2}\,\left (1
-{4 m^2 \over q^2}\right )+ \frac{\alpha_s(\mu^2)}{4\pi} C_F \Bigg \{
     \Bigg(-32+{32 m^2 \over q^2}  
\nonumber \\[2ex]
&&  + {4 q^6 \over \mpq^3}-{10q^4 \over \mpq^2}+{25 q^2 \over \mpq} \Bigg) 
\lnsqt
\nonumber \\[2ex]
&&    +{1 \over \sqrq} \Bigg (28 m^2 -7 q^2\Bigg ) \ln(t) 
-2 -{2 q^4 \over \mpq^2} 
\nonumber \\[2ex]
&& -{8 m^2 \over q^2 }  +{4 q^2 \over \mpq}  -\left({16 m^4 \over q^2 } 
-16 m^2 +{q^2 \over 2} \right) {\cal J} \Bigg \} \Bigg ]
\nonumber\\[2ex]
& \!\!\mathop{\mbox{=}}\limits_{\vphantom{\frac{A}{A}} -q^2 \gg m^2}\!\!&
{\it v}_q^{V_1}{\it v}_q^{V_2} \Bigg [
      \frac{1}{2} + \frac{\alpha_s(\mu^2)}{4\pi}C_F \Bigg \{-1
- \moq \Bigg(\frac{37}{9} +\frac{10}{3}\ln\moq\Bigg) \Bigg \}\Bigg ]
\nonumber \\
&&  + {\it a}_q^{V_1}{\it a}_q^{V_2} \Bigg [\frac{1}{2} + \frac{\alpha_s(\mu^2)}
{4\pi} C_F \Bigg \{ -1- \moq \Bigg(\frac{91}{9}-\frac{26}{3} \ln\moq \Bigg) 
\Bigg \} \Bigg]\,,
\\[2ex]
\label{eq:4.22}
{\cal C}_{2,q}^{(1)}&=&\int_0^1 {dz \over z} \hat {\cal F}_2(z,q^2,m^2) = 
{\it v}_q^{V_1}{\it v}_q^{V_2}+{\it a}_q^{V_1}{\it a}_q^{V_2}
\\[2ex]
\label{eq:4.23}
{\cal C}_{3,q}&=&\int_0^1 dz \hat {\cal F}_3(z,q^2,m^2)=  
\left ({\it v}_q^{V_1}{\it a}_q^{V_2} +{\it a}_q^{V_1}{\it v}_q^{V_2} \right )
\Bigg [ 1+ \frac{\alpha_s(\mu^2)}{4\pi} C_F \Bigg \{
        \Bigg(-16 +{2 q^2 \over m^2} 
\nonumber \\[2ex]
&& +{2 q^4 \over \mpq^2}-{2 q^2 \over \mpq} \Bigg) \lnsqt 
+{1 \over \sqrq} \left ( 24 m^2 +{ q^4 \over m^2} \right. 
\nonumber \\[2ex]
&& \left. -10 q^2\right)\ln(t) -4 -{ q^2 \over \mpq} + 8 m^2 {\cal J} \Bigg \} 
\Bigg ]
\nonumber\\[2ex]
&\!\!\mathop{\mbox{=}}\limits_{\vphantom{\frac{A}{A}} -q^2 \gg m^2}\!\!&
\left ({\it v}_q^{V_1}{\it a}_q^{V_2} +{\it a}_q^{V_1}{\it v}_q^{V_2} \right )
\Bigg [1 +  \frac{\alpha_s(\mu^2)}{4\pi}C_F \Bigg \{ -3+\moq \Bigg( -4
\nonumber\\[2ex]
&& -3 \ln\moq \Bigg) \Bigg \} \Bigg]\,,
\\[2ex]
\label{eq:4.24}
\delta {\cal C}_{1,q}&=&\left (\Gamma_q^5\right )^{-1}\int_0^1 dz 
\hat g_1(z,q^2,m^2) = {\it v}_q^{V_1}{\it v}_q^{V_2} \Bigg[ \frac{1}{2} +
\frac{\alpha_s(\mu^2)}{4\pi} C_F \Bigg \{ \Bigg(-8 -{q^2 \over m^2} 
\nonumber\\[2ex]
&&  +{q^2 \over 2 \mpq}\Bigg) \lnsqt +{1 \over \sqrq} \left(14 m^2 
-{q^4 \over 2 m^2} -{5 q^2 \over 2} \right) \ln(t)
\nonumber \\[2ex]
&&     -1+\left(4 m^2 +{q^2 \over 4} \right) {\cal J} \Bigg \}\Bigg]
\nonumber \\[2ex]
&& +{\it a}_q^{V_1}{\it a}_q^{V_2} \Bigg[ \frac {1}{2} + \frac{\alpha_s(\mu^2)}{4\pi} C_F \Bigg \{
     \Bigg(-{q^2 \over m^2} + {2 q^4 \over \mpq^2}
           +{q^2 \over 2 \mpq} \Bigg) \lnsqt
\nonumber \\[2ex]
&&   +{1 \over \sqrq} \left(-2 m^2 - {q^4 \over 2 m^2} 
        +{5 q^2 \over 2}\right) \ln(t)
\nonumber \\[2ex]
&& -{ q^2 \over \mpq} + {q^2 \over 4} {\cal J} \Bigg \}\Bigg]
\nonumber\\[2ex]
&\!\!\mathop{\mbox{=}}\limits_{\vphantom{\frac{A}{A}} -q^2 \gg m^2}\!\!&
{\it v}_q^{V_1}{\it v}_q^{V_2}
\Bigg [ \frac{1}{2} + \frac{\alpha_s(\mu^2)}{4\pi}C_F \Bigg \{
        -{3 \over 2} + \moq \Bigg(\frac{5}{9} -\frac{11}{6} 
\ln\moq\Bigg) \Bigg \}
       \Bigg ]
\nonumber\\[2ex]
&& +{\it a}_q^{V_1}{\it a}_q^{V_2} \Bigg[\frac{1}{2} + \frac{\alpha_s(\mu^2)}
{4\pi}C_F \Bigg \{ -{3 \over 2} - \moq \Bigg(\frac{22}{9} +\frac{11}{6} 
\ln\moq \Bigg) \Bigg \} \Bigg]\,,
\\[2ex]
\label{eq:4.25}
\delta {\cal C}_{4,q}&=&\left (\Gamma_q^5\right )^{-1}
\int_0^1 {dz \over z}\hat g_4(z,q^2,m^2) =
- \left ({\it v}_q^{V_1}{\it a}_q^{V_2} + {\it a}_q^{V_1}{\it v}_q^{V_2}\right )
\,,
\\[2ex]
\label{eq:4.26}
\delta {\cal C}_{5,q}&=&\left (\Gamma_q^5\right )^{-1}\int_0^1 dz 
\hat g_5(z,q^2,m^2) = \left ({\it v}_q^{V_1}{\it a}_q^{V_2}
+{\it a}_q^{V_1}{\it v}_q^{V_2}
\right ) \Bigg [-\frac{1}{2} + \frac{\alpha_s(\mu^2)}{4\pi} C_F \Bigg \{
\nonumber \\[2ex]
&&   \Bigg(8 - { q^4 \over \mpq^2} +{3 q^2 \over 2\mpq} \Bigg) \lnsqt 
+{1 \over \sqrq} \Bigg(-14 m^2   
\nonumber \\[2ex]
&& +\frac{7}{2} q^2\Bigg) \ln(t)+2 +{q^2 \over 2\mpq} -\left(4 m^2 
+{3 q^2 \over 4} \right) {\cal J} \Bigg \} \Bigg]
\nonumber\\[2ex]
&\!\!\mathop{\mbox{=}}\limits_{\vphantom{\frac{A}{A}} -q^2 \gg m^2}\!\!&
\left ({\it v}_q^{V_1}{\it a}_q^{V_2} + {\it a}_q^{V_1}{\it v}_q^{V_2}
\right )\Bigg[-\frac{1}{2}+ \frac{\alpha_s(\mu^2)}{4\pi}C_F
\Bigg \{ 1+\moq \Bigg( {4 \over 3}
\nonumber\\[2ex]
&& +2 \ln \moq \Bigg) \Bigg \} \Bigg ]\,,
\end{eqnarray}
where the integral ${\cal J}$ is defined as
\begin{eqnarray}
\label{eq:4.27}
{\cal J}&=&\int_{m^2}^{\infty} d\hat s {1 \over \lsq \smq}\ln\left( 
{\hat s + m^2 - q^2 - \sqrt{\lam} \over
        \hat s + m^2 - q^2 + \sqrt{\lam} }\right)
\nonumber \\[2ex]
&=&        { 1\over 2 m \sqrt{-q^2}} \Bigg[- 4 {\cal L}i_2\left(1-\sqt\right) 
           - 4 {\cal L}i_2\left(-\sqt\right)
           +2 {\cal L}i_2\left(-{t-1 \over \sqt}\right) 
\nonumber \\[2ex]
&&       - 2 {\cal L}i_2\left(1-{\sqt \over t-1}\right)  
           -2 \ln(t) \ln(1+\sqt) +{1 \over 4 } \ln^2(t) -\ln^2(t-1) 
\nonumber \\[2ex]
&&  -\ln(t) \ln(t+\sqt-1) +2 \ln(t-1) \ln(t+\sqt-1)-4 \zeta(2) \Bigg]\,.
\end{eqnarray}
The result for ${\cal C}_{3,q}$ in Eq. (\ref{eq:4.23}) and the vector-vector
part of $\delta {\cal C}_{1,q}$ in Eq. (\ref{eq:4.24}) is already presented
in \cite{teve}.
The reason that the sum rules in Eqs. (\ref{eq:4.22}), (\ref{eq:3.25}) do not 
receive order $\alpha_s$ corrections is a consequence of the ETC algebra
presented in  Eq. (\ref{eq:2.5}). This is because the order $\alpha_s$
coefficient functions are the same for charged current as well as neutral 
current interactions. However in next order this property does not hold
anymore. From \cite{zn1}, \cite{zn2} one can infer that the coefficient
functions are different due to the sign of the contributions coming from
processes with equal quarks in the final state given by the reaction
$V + q\rightarrow q + q +\bar q$. It turns out that sum rules corresponding
to charged current interactions interactions do not receive order 
$\alpha_s^2$ corrections which is the case for the Adler sum rule in
Eq. (\ref{eq:2.24}) and its polarized analogue in Eq. (\ref{eq:2.26}).
Notice that this statement is verified for $m=0$ but it has to hold for
$m\not =0$ as well. However for the neutral current sum rules like
the Gottfried sum rule in Eq. (\ref{eq:3.57}) one obtains order $\alpha_s^2$
corrections which even contain scaling violating terms of the type
$\ln -q^2/m^2$ which have to be absorbed in $\Gamma_q$ (see also below
Eq. (\ref{eq:3.57})). Therefore one should be careful in drawing to premature
conclusions about the vanishing of the order $\alpha_s$ corrections to
sum rules. From the computations above we also observe that if a
sum rule gets order $\alpha_s$ corrections it also receives power corrections
and vice versa. These power corrections are of the type $m^2/q^2$ and they
are a signature of higher twist contributions.
Further in the limit $m \rightarrow 0$ we have the following relations
which hold up to order $\alpha_s^2$ (see \cite{latk}, \cite{lave}, \cite{zn1},
\cite{zn2})
\begin{eqnarray}
\label{eq:4.28}
{\cal C}_{1,q}=\delta {\cal C}_{5,q}\,, \qquad
{\cal C}_{3,q}=\delta {\cal C}_{1,q}\,.
\end{eqnarray}
These relations are broken in order $\alpha_s^3$. In \cite{lave} one has 
found that in this order ${\cal C}_{3,q}\not =\delta {\cal C}_{1,q}$.
Probably this also holds for ${\cal C}_{1,q},~\delta {\cal C}_{5,q}$
although the third order result for the latter is not known yet.
We also want to emphasize that the correct expressions for the coefficient
functions are only obtained if the renormalization of the axial-vector current 
given by $\Gamma_q^5$ is taken into account. This is sometimes forgotten
in the literature (see e.g. \cite{teve}, \cite{ais}). Hence it is incorrect
to conclude that in the limit $-q^2 \gg m^2$ the polarized coefficient 
functions $\delta {\cal C}_{i,q}$ depend on the choice of the mass i.e.
$m=0$ versus $m\not =0$. In the case $m=0$, where $\Gamma_q^5=1$ (see
the remark below Eq. (\ref{eq:4.5})), one can identify the first moments
of the partonic structure function and the coefficient functions but
for $m\not =0$ this is not allowed. Another remark we want to make is
that one should be careful with n-dimensional regularization when the 
$\gamma_5$ matrix is present. There exist prescriptions which break the 
conservation of the non-singlet axial-vector
current. To restore the Ward identities one needs an additional renormalization
constant which is the analogue of $\Gamma_q^5$ introduced in the equations 
above. In this case the renormalization constant
can even become infinite (see e.g. \cite{lave}).

Next we discuss the sum rules, derived at the end of section 3, which also 
involve leading twist three contributions. The first
one is the Burkhardt-Cottingham sum rule \cite{buco} in Eq. (\ref{eq:3.58})
Using Eqs. (\ref{eq:B5}) and (\ref{eq:B16}) one obtains up to order
$\alpha_s$
\begin{eqnarray}
\label{eq:4.29}
\int_0^1 dz\, \hat g_2(z,q^2,m^2) = 0\,.
\end{eqnarray}
This result was already checked for the vector-vector part in the
context of QED in \cite{tdm} (see also \cite{mene}, \cite{alnr}). In this paper
we have shown that it also holds for the axial-vector-axial-vector
part. The second sum rule presented in Eq. (\ref{eq:3.59}) also yields zero up 
to next to leading order
\begin{eqnarray}
\label{eq:4.30}
\int_0^1dz\,\left (\hat g_4(z,q^2,m^2) -\hat g_3(z,q^2,m^2)\right )=0\,.
\end{eqnarray}
Therefore the result obtained for the Burkhardt-Cottingham sum rule is
not unique. On the other hand the sum rule in (see Eq. (\ref{eq:3.60})) is
non-vanishing since it receives order $\alpha_s$ contributions.
\begin{eqnarray}
\label{eq:4.31}
&& \int_0^1dz\,\left (2\,z\,\hat g_5(z,q^2,m^2) -\hat g_3(z,q^2,m^2)\right )=
\nonumber\\[2ex]
&& \left ({\it v}_q^{V_1}{\it a}_q^{V_2} +{\it a}_q^{V_1}{\it v}_q^{V_2}\right )
\, \frac{\alpha_s(\mu^2)}{4\pi}C_F \,\Bigg[ \Bigg(-{29 q^2 \over 2 m^2}
 - {17 q^2 \over 2\mpq} +{7 q^4 \over \mpq^2} -{4 q^6 \over \mpq^3} \Bigg) 
\nonumber\\[2ex]
&&\times \lnsqt +{1 \over \sqrq }\Bigg(-{13 q^4 \over 2 m^2} +12 m^2 
+23 q^2 \Bigg) \ln(t) -8-{3 q^2 \over 2 m^2} -{5 q^2 \over 2 \mpq} 
\nonumber\\[2ex]
&&+{2 q^4 \over \mpq^2}
+\Bigg({3 q^4 \over 4 m^2} +{23 q^2 \over 2} \Bigg) {\cal J} \Bigg]
\nonumber\\[2ex]
&& \!\!\mathop{\mbox{=}}\limits_{\vphantom{\frac{A}{A}} -q^2 \gg m^2}\!\!
 \left ({\it v}_q^{V_1}{\it a}_q^{V_2} +{\it a}_q^{V_1}{\it v}_q^{V_2}\right )
\, \frac{\alpha_s(\mu^2)}{4\pi}C_F \,
\Bigg[{4 \over 3} + \left({m^2 \over -q^2}\right) \Bigg({1124 \over 225}
+{76 \over 15} \ln\left( {m^2 \over -q^2} \right) \Bigg) \Bigg]\,.
\end{eqnarray}
Finally we also present the lowest order corrections to the sum rules given
in Eqs. (\ref{eq:3.61}), (\ref{eq:3.62}). The former receives higher order
corrections and it reads
\begin{eqnarray}
\label{eq:4.32}
&& (\Gamma_q^5)^{-1}\,\int_0^1\frac{dz}{z}\,\left (\hat g_3(z,q^2,m^2)
+2\,z\,\hat g_9(z,q^2,m^2) \right )=
\nonumber\\[2ex]
&& -2\,
\left ({\it v}_q^{V_1}{\it a}_q^{V_2} +{\it a}_q^{V_1}{\it v}_q^{V_2}\right )
\Bigg [1 +  \frac{\alpha_s(\mu^2)}{4\pi}C_F \Bigg \{ 
\Bigg (-8 + \frac{2q^4}{\mpq^2} \Bigg )\lnsqt 
\nonumber\\[2ex]
&& +{1 \over \sqrq} \Bigg(8 m^2 -3 q^2\Bigg) \ln(t) + {m^2 \over \mpq} 
+4 m^2 {\cal J} \Bigg \} \Bigg]
\nonumber\\[2ex]
&&\mathop{\mbox{=}}\limits_{\vphantom{\frac{A}{A}} -q^2 \gg m^2}
-2\,
\left ({\it v}_q^{V_1}{\it a}_q^{V_2} +{\it a}_q^{V_1}{\it v}_q^{V_2}\right )
\Bigg [1 +  \frac{\alpha_s(\mu^2)}{4\pi}C_F \Bigg \{ \frac{3m^2}{q^2} \Bigg \}
\Bigg ]\,.
\end{eqnarray}
On the other hand the sum rule in Eq. (\ref{eq:3.62}) 
does not get any corrections and it reads
\begin{eqnarray}
\label{eq:4.33}
(\Gamma_q^5)^{-1}\,\int_0^1\frac{dz}{z}\,\left (\hat g_3(z,q^2,m^2)
-2\,z\,\hat g_8(x,q^2,m^2)
\right )= -2\, \left ({\it v}_q^{V_1}{\it a}_q^{V_2} +{\it a}_q^{V_1}
{\it v}_q^{V_2}\right )\,.
\end{eqnarray}
One observes a similarity in behaviour between the sum rules in 
Eqs. (\ref{eq:4.25}) and Eqs. (\ref{eq:4.33}) which also holds for 
Eqs. (\ref{eq:4.26}) and (\ref{eq:4.32}). The reason for
the vanishing of the first order contributions to the sum rules in Eqs.
(\ref{eq:4.29}), (\ref{eq:4.30}), (\ref{eq:4.33})
is rather obscure. Probably it is the consequence of a super-convergence
relation. Suppose the forward Compton scattering amplitude 
$\Delta T(\nu,q^2,m^2)$ ($\nu=p\cdot q/m$) behaves asymptotically like 
$\nu^{-1-a}$ with $a>0$
then it satisfies the unsubtracted dispersion relation
\begin{eqnarray}
\label{eq:4.34}
\Delta T(\nu,q^2,m^2)&=&\frac{1}{\pi}\int_{-q^2/2m}^{\infty}d\nu'\,
\frac{\Delta F(\nu',q^2,m^2)}{\nu'-\nu} \,,
\nonumber\\[2ex]
 \Delta F(\nu,q^2,m^2)&=&
{\rm Im} \, \Delta T(\nu,q^2,m^2)\,,
\end{eqnarray}
where the symbol $\Delta$ denotes a combination of amplitudes or structure
functions. Taking the limit $\nu \rightarrow \infty$ provides us
with the super convergence relation
\begin{eqnarray}
\label{eq:4.35}
\int_0^1\frac{dx}{x^2}\,\Delta F(x,q^2,m^2)=0\,,
\end{eqnarray}
In the derivation of the formula above we have used the substitution 
$\nu=-q^2/2mx$. If we take
$\Delta F=\hat g_2/\nu^2$ and $\Delta F=(\hat g_4-\hat g_3)/\nu^2$, 
Eqs. (\ref{eq:4.29}) and (\ref{eq:4.30}) follow automatically. 
Apparently $\Delta T$ satisfies
the requirement for the asymptotic behaviour given above.
In a similar way one can take $\Delta F=(\hat g_3^{(1)}
-2\,x\,\hat g_8^{(1)})/\nu$ leading to
the vanishing of the order $\alpha_s$ correction in Eqs. (\ref{eq:4.33}).
However there is no general argument in quantum field theory that this 
behaviour persists in all orders. In fact the non-zero contributions in
Eqs. (\ref{eq:4.31}), (\ref{eq:4.32}) already indicate that the reasons
behind the existence of the super convergence relation are obscure. 
The only argument is given by the ETC algebra in Eq. (\ref{eq:2.5})
which predicts that the Adler sum rule  Eq. (\ref{eq:2.24}) and its 
polarized analogue in  Eq. (\ref{eq:2.26}) do not receive any higher order
corrections.

Summarizing our findings we conclude that there exist two type of
sum rules which, because of their origin, are called fundamental and 
non-fundamental sum rules. The former are related to expectation
values of conserved currents or partially conserved axial-vector currents
which are sandwiched between hadronic
states. The ones which originate from the equal time current (ETC)
algebra are independent of the nature of the (axial-) vector currents
and they do not receive higher order contributions in perturbation theory 
or power corrections. The other ones only hold for fermionic currents
characteristic of QCD and they receive QCD as well as power corrections.
Both types provide us with stringent tests for QCD. The non-fundamental
sum rules which originate from the leading twist parton model are unstable
against higher order QCD corrections at least from second order in $\alpha_s$
onwards. They receive scaling violating terms which indicates that they
are heavily broken. Also non-fundamental are those sum rules which appear
in polarized scattering and contain twist three contributions. One of
them is the Burkhardt-Cottingham sum rule, derived for the polarized structure 
function $g_2$, which vanishes up to first order
for the vector current as well as axial-vector current. However this
sum rule is not unique since this property also holds for other structure
functions. In quantum field theory the origin of this property is 
still obscure and we do not know whether the corrections to these sum 
rules will vanish in higher order.\\[5mm]
\noindent
ACKNOWLEDGMENTS\\[3mm]
The work of W.L. van Neerven was supported
by the EC network `QCD and Particle Structure' under contract
No.~FMRX--CT98--0194.


\appendix
\mysection{Appendix A}
In this section we present the sum rules in the case of four flavours
i.e. $SU(4)_F$. For this group the charged current in Eq. (\ref{eq:2.16})
becomes equal to
\begin{eqnarray}
\label{eq.A1}
J_{\pm}^{\mu}(y)&=&\Big (V_{1\pm i2}^{\mu}(y)-A_{1\pm i2}^{\mu}(y) +
V_{13\mp i14}^{\mu}(y)-A_{13\mp i14}^{\mu}(y)\Big ) \cos \theta_c
\nonumber\\[2ex]
&& +\Big (V_{4 \pm i5}^{\mu}(y)- A_{4 \pm i5}^{\mu}(y)-
V_{11\mp i12}^{\mu}(y)+A_{11\mp i12}^{\mu}(y) \Big ) \sin \theta_c\,.
\end{eqnarray}
Using the infinite momentum frame technique the Adler sum rule equals
\begin{eqnarray}
\label{eq.A2}
&& \int_0^1 \frac{d~x}{x}\left (F_2^{\rm W^-N}(x,Q^2) - F_2^{\rm W^+N}(x,Q^2)
 \right )= 4\,I_3^N+2\,S^N+2\,C^N\,,
\end{eqnarray}
and for polarized scattering we obtain
\begin{eqnarray}
\label{eq.A3}
&& \int_0^1  \frac{d~x}{x} \left (g_4^{\rm W^-N}(x,Q^2) - g_4^{\rm W^+N}(x,Q^2)
 \right )= -4\,I_3^N(F'+D')-2\,F'
\nonumber\\[2ex]
&& +\frac{2}{3}(D'+2\,E)\,.
\end{eqnarray}
In the formulae above the vector charges (see  Eq. (\ref{eq:2.25}) are given by
\begin{eqnarray}
\label{eq.A4}
&& \Gamma_3^N=2\,I_3^N \,,\quad \Gamma_8^N=\sqrt 3\,(B^N+S^N-\frac{1}{3}\,C^N)
\,,\quad \Gamma_{15}^N= \frac{1}{2}\sqrt 6\,(B^N-\frac{4}{3}\,C^N)\,,
\nonumber\\[2ex]
&& \Gamma_0^N=3\,B^N\,,
\end{eqnarray}
where $C^N$ denotes the quantum number for charm.
The axial-vector charges (see Eq. (\ref{eq:2.27})) are equal to
\begin{eqnarray}
\label{eq.A5}
\Gamma_3^{5,N}=2\,I_3^N (F'+D')\,, \qquad \Gamma_8^{5,N}=\frac{1}{3}\sqrt 3\,
(3\,F'-D')\,, \qquad \Gamma_{15}^{5,N}=\frac{1}{3}\sqrt 6\,E\,.
\end{eqnarray}
Besides $E$, which emerges from the expectation value $\langle N(p,s) |
A_{15}^{\mu}(0)|N(p,s) \langle $ when the flavour symmetry is given by $SU(4)$,
the numerical values for $F$ and $D$ change from those given for
$SU(3)_F$. Therefore we have indicated them by a prime because the
original values were obtained by assuming a $SU(3)_F$ symmetry.
In the case of the proton (p) and the neutron (n) the quantum numbers above
are given by
\begin{eqnarray}
\label{eq.A6}
I_3^p=\frac{1}{2} \,,\qquad B^p=1\,, \qquad S^p=0 \,,\qquad C^p=0\,,
\nonumber\\[2ex]
I_3^n=-\frac{1}{2} \,,\qquad B^n=1 \,,\qquad S^n=0 \qquad C^n=0\,.
\end{eqnarray}
Following the same procedure as in the case of three flavours 
one can derive from the light-cone current algebra in Eqs. (\ref{eq:3.1}),
(\ref{eq:3.2}) the charged current structure functions for four flavours
\begin{eqnarray}
\label{eq.A7}
F_2^{\rm W^+ N}(x)/x&=&2\,\bar V_0^1(x)-2\,V_3^1(x)
-\frac{2}{3}\sqrt 3\,V_8^1(x) +\frac{2}{3}\sqrt 6\,V_{15}^1(x)\,,
\\[2ex]
\label{eq.A8}
F_3^{\rm W^+ N}(x)/x&=&2\,V_0^1(x)-2\,\bar V_3^1(x)
-\frac{2}{3}\sqrt 3\,\bar V_8^1(x) +\frac{2}{3}\sqrt 6\,\bar V_{15}^1(x)\,,
\\[2ex]
\label{eq.A9}
g_1^{\rm W^+ N}(x)&=&\left (A_0^1(x)+\frac{\partial A_0^2(x)}{\partial x}
 \right ) -\left (\bar A_3^1(x)+\frac{\partial \bar A_3^2(x)}{\partial x}
\right ) -\frac{1}{3}\sqrt 3\,\left (\bar A_8^1(x)+\frac{\partial \bar A_8^2(x)}
{\partial x} \right )
\nonumber\\[2ex]
&& +\frac{1}{3}\sqrt 6\, \left ( \bar A_{15}^1(x)
+\frac{\partial \bar A_{15}^2(x)}{\partial x} \right )\,,
\\[2ex]
\label{eq.A10}
g_4^{\rm W^+ N}(x)/x&=&-2\, \left (\bar A_0^1(x)
+\frac{\partial \bar A_0^2(x)}{\partial x} \right )
+2\, \left (A_3^1(x)+\frac{\partial A_3^2(x)}{\partial x} \right )
\nonumber\\[2ex]
&& +\frac{2}{3}\sqrt 3\,\left ( A_8^1(x)+\frac{\partial A_8^2(x)}
{\partial x} \right )
-\frac{2}{3}\sqrt 6\,\left ( A_{15}^1(x)+\frac{\partial A_{15}^2(x)}
{\partial x} \right )\,.
\end{eqnarray}
The structure functions for $F_i^{\rm W^- p}$ and $g_i^{\rm W^- p}$ are
obtained in the same way as mentioned below Eqs. 
(\ref{eq:3.16})-(\ref{eq:3.19}).
The sum rules are given by
\begin{eqnarray}
\label{eq.A11}
&& \mbox{unpolarized Bjorken sum rule}
\nonumber\\[2ex]
&&\int_0^1 d~x\left (F_1^{\rm W^-N}(x,Q^2) - F_1^{\rm W^+N}(x,Q^2) \right )
= 2\,I_3^N+S^N+C^N\,,
\\[2ex]
&& \mbox{Gross Llewellyn Smith sum rule}
\label{eq.A12}
\nonumber\\[2ex]
&&\int_0^1 d~x\left (F_3^{\rm W^-N}(x,Q^2) + F_3^{\rm W^+N}(x,Q^2) \right )
= 6\,B^N\,,
\\[2ex]
\label{eq.A13}
&& \int_0^1  d~x \left (g_5^{\rm W^-N}(x,Q^2) - g_5^{\rm W^+N}(x,Q^2) \right )
=-2\,I_3^N\,(F'+D')-F'
\nonumber\\[2ex]
&& \hspace*{7cm} +\frac{1}{3}\,(D'+2\,E)\,.
\end{eqnarray}
The neutral current structure functions are given by the same expressions as
derived for $SU(3)_F$ in Eqs. (\ref{eq:3.34})-(\ref{eq:3.43}) except that one 
has to replace the following equations. The neutral currents in Eq.
(\ref{eq:3.29}) are now given by
\begin{eqnarray}
\label{eq.A14}
V_{\gamma}^{\mu}(y) &\equiv& V_3^{\mu}(y)+\frac{1}{3}\sqrt 3\,V_8^{\mu}(y)
-\frac{1}{3}\sqrt 6\,V_{15}^{\mu}(y)+\frac{1}{3}\,V_0^{\mu}(y)\,,
\nonumber\\[2ex]
A_{\gamma}^{\mu}(y)&\equiv& A_3^{\mu}(y)+\frac{1}{3}\sqrt 3\,A_8^{\mu}(y)
-\frac{1}{3}\sqrt 6\,A_{15}^{\mu}(y)+\frac{1}{3}\,A_0^{\mu}(y)\,,
\end{eqnarray}
and the formulae in Eq. (\ref{eq:3.33}) are replaced by
\begin{eqnarray}
\label{eq.A15}
V_{\gamma}(x)&=&V_3^1(x)+\frac{1}{3}\sqrt 3V_8^1(x)-\frac{1}{3}\sqrt 6V_8^1(x)
+\frac{1}{3}V_0^1(x)\,,
\nonumber\\[2ex]
A_{\gamma}(x)&=&A_3^1(x)+\frac{\partial A_3^2(x)}{\partial x}
+\frac{1}{3}\sqrt 3\left (A_8^1(x)+\frac{\partial A_8^2(x)} {\partial x}\right )
-\frac{1}{3}\sqrt 6\left (A_{15}^1(x)+\frac{\partial A_{15}^2(x)}{\partial x}
\right )
\nonumber\\[2ex]
&&+\frac{1}{3}\left (A_0^1(x)+\frac{\partial A_0^2(x)} {\partial x} \right )\,,
\nonumber\\[2ex]
V_{\gamma \gamma}(x)&=&\frac{5}{9}\,V_0^1(x)+\frac{1}{3}\,V_3^1(x)
+\frac{1}{9}\sqrt 3\,V_8^1(x)-\frac{1}{9}\sqrt 6\,V_{15}^1(x)\,,
\nonumber\\[2ex]
A_{\gamma \gamma}(x)&=&\frac{5}{9}\left (A_0^1(x)+\frac{\partial A_0^2(x)}
{\partial x}\right )+\frac{1}{3}\left (A_3^1(x)+\frac{\partial A_3^2(x)}
{\partial x}\right )
\nonumber\\[2ex]
&& +\frac{1}{9}\sqrt 3\left (A_8^1(x)+\frac{\partial A_8^2(x)} {\partial x}
\right )-\frac{1}{9}\sqrt 6 \left (A_{15}^1(x)+\frac{\partial A_{15}^2(x)}
{\partial x}\right )\,.
\end{eqnarray}
The sum rules in Eqs. (\ref{eq:3.44}), (\ref{eq:3.45}) become for $SU(4)_F$
\begin{eqnarray}
\label{eq.A16}
\int_0^1 dx\,F_3^{\rm Z N}(x,Q^2) &=& \frac{3}{2}\,(1-2\,s^2)\,B^N
-\frac{1}{3}\,s^2\,(2\,I_3^N+S^N+C^N)\,,
\\[2ex]
\label{eq.A17}
\int_0^1 dx\,F_3^{\rm \gamma Z~N}(x,Q^2)&=&\frac{3}{2}\,B^N+
+\frac{1}{6}\,(2\,I_3^N+S^N+C^N)\,.
\end{eqnarray}
The sum rules for the longitudinal structure function $g_1$ have the same
form as in Eqs. (\ref{eq:3.46})-(\ref{eq:3.48}). However the shorthand 
notations in Eq. (\ref{eq:3.49}) read
\begin{eqnarray}
\label{eq.A18}
f^{\rm NS}&=&\frac{1}{6}\,I_3^N\,(F'+D')+\frac{1}{12}\,F'-\frac{1}{36}(D'+2\,E)
\,,
\nonumber\\[2ex]
f_{\gamma}^{\rm S}&=&\frac{5}{18}\,\Gamma_0^{5,N}\,,
\nonumber\\[2ex]
f_{\rm Z}^{\rm S}&=&(\frac{1}{2}-s^2+\frac{10}{9}\,s^4)\,\Gamma_0^{5,N}\,,
\nonumber\\[2ex]
f_{\gamma Z}^{\rm S}&=&(\frac{1}{2}-\frac{10}{9}\,s^2)\,\Gamma_0^{5,N}\,.
\end{eqnarray}
The bilocal operator matrix elements are expressed into the parton densities
in the same way as presented in 
Eqs. (\ref{eq:3.54}), (\ref{eq:3.56}) except that one has additional 
matrix elements
due to the generator $\lambda_{15}$ appearing in $SU(4)$. The latter are given 
by
\begin{eqnarray}
\label{eq.A19}
V_{15}^1(x)&=&\frac{1}{2\sqrt 6}\Big (V_u(x)+V_d(x)+V_s(x)-3\,V_c(x)\Big )\,,
\nonumber\\[2ex]
\bar V_{15}^1(x)&=&\frac{1}{2\sqrt 6}\Big (\Delta_u(x)+\Delta_d(x)+\Delta_s(x)
-3\,\Delta_c(x)\Big )\,,
\nonumber\\[2ex]
A_{15}^1(x)+\frac{\partial A_{15}^2(x)}{\partial x}&=&\frac{1}{2\sqrt 6}
\Big (\Delta_{\delta u}(x)+\Delta_{\delta d}(x)+\Delta_{\delta s}(x)-3\,
\Delta_{\delta c}(x)\Big )\,,
\nonumber\\[2ex]
\bar A_{15}^1(x)+\frac{\partial \bar A_{15}^2(x)}{\partial x}&=&
\frac{1}{2\sqrt 6}\Big (V_{\delta u}(x)+V_{\delta d}(x)+V_{\delta s}(x)
-3\,V_{\delta c}(x)\Big )\,.
\end{eqnarray}
The definitions for the non-singlet and singlet quark densities are changed 
into
\begin{eqnarray}
\label{eq.A20}
&& V_q(x)=q(x)-\bar q(x)\,, \qquad \Delta_q(x)=q(x)+\bar q(x)-\frac{1}{4}
\Sigma(x) \,,
\nonumber\\[2ex]
&&\Sigma(x)=\sum_{q=u,d,s,c} q(x)+\bar q(x) \,,
\end{eqnarray}
with similar notations for the polarized quark densities.

\mysection{Appendix B}
In this appendix we only present the results for the structure functions
which are observed in the cross section when the lepton masses are neglected.
Therefore we omit the results for $F_i$ ($i=4,5$) and $g_i$ ($i=6-9$).
The order $\alpha_s$ contributions to the integral in Eq. (\ref{eq:4.20}),
coming from the sum of soft and virtual gluon corrections, are given by
\begin{eqnarray}
\label{eq:B1}
{\cal T}_{1,q}^{\rm S+V}(z)\!\!&=&\!\!C_F\Bigg[
                   {\it v}_q^{V_1} {\it v}_q^{V_2} \cvs 
                   + {\it a}_q^{V_1} {\it a}_q^{V_2}
                   \left(1-{4 m^2 \over q^2} \right)
                  \Big(\cvs 
                     +2 {\cal R}_2 \Big) 
                  \Bigg]\,,
\nonumber\\
\\[2ex]
\label{eq:B2}
{\cal T}_{2,q}^{\rm S+V}(z)\!\!&=&\!\! 2 C_F\Bigg[
                    {\it v}_q^{V_1} {\it v}_q^{V_2} \Big(\cvs+{\cal R}_2\Big) 
                    +{\it a}_q^{V_1} {\it a}_q^{V_2} \Big(\cvs
                   +2 {\cal R}_2\Big) \Bigg]\,,
\\[2ex]
\label{eq:B3}
{\cal T}_{3,q}^{\rm S+V}(z)\!\!&=&\!\! 2 C_F\Bigg[\Big ({\it v}_q^{V_1} 
{\it a}_q^{V_2}+{\it a}_q^{V_1} {\it v}_q^{V_2}\Big ) \Big(\cvs+{\cal R}_2\Big) 
                    \Bigg]\,,
\\[2ex]
\label{eq:B4}
\delta {\cal T}_{1,q}^{\rm S+V}(z)\!\!&=&\!\!  C_F\Bigg[
     {\it v}_q^{V_1} {\it v}_q^{V_2} \Big(\cvs+{1 \over 2} {\cal R}_2\Big) 
 +   {\it a}_q^{V_1} {\it a}_q^{V_2} \Big(\cvs+2 {\cal R}_2\Big) \Bigg]\,,
\\[2ex]
\label{eq:B5}
\delta {\cal T}_{2,q}^{\rm S+V}(z)\!\!&=&\!\!-{q^2 \over 8 m^2}  C_F\Bigg[
      {\it v}_q^{V_1} {\it v}_q^{V_2}               {\cal R}_2 
                    +{\it a}_q^{V_1} {\it a}_q^{V_2} {\cal R}_3 
                  \Bigg]\,,
\\[2ex]
\label{eq:B6}
\delta {\cal T}_{3,q}^{\rm S+V}(z)\!\!&=&\!\!- C_F\Bigg[
  \Big ({\it v}_q^{V_1} {\it a}_q^{V_2}+{\it a}_q^{V_1} {\it v}_q^{V_2}\Big ) 
             \Big(2 \cvs +3 {\cal R}_2-{q^2 \over 4 m^2 } {\cal R}_2\Big) 
                  \Bigg]\,,
\\[2ex]
\label{eq:B7}
\delta {\cal T}_{4,q}^{\rm S+V}(z)\!\!&=&\!\! - C_F\Bigg[
\Big ({\it v}_q^{V_1} {\it a}_q^{V_2}+{\it a}_q^{V_1} {\it v}_q^{V_2}\Big ) 
                  \Big(2 \cvs +3 {\cal R}_2\Big) 
                  \Bigg]\,,
\\[2ex]
\label{eq:B8}
\delta {\cal T}_{5,q}^{\rm S+V}(z)\!\!&=&\!\! - C_F\Bigg[
\Big ({\it v}_q^{V_1} {\it a}_q^{V_2}+{\it a}_q^{V_1} {\it v}_q^{V_2}\Big )
           \Big(\cvs +{\cal R}_2\Big) \Bigg]\,,
\end{eqnarray}
with the definition
\begin{eqnarray}
\label{eq:B9}
\cvs &=& S^{\rm SOFT}+{\cal R}_1= {\alpha_s \over 4 \pi} C_F \Bigg[
             \left({4 m^2 -2 q^2 \over \sqrq} \ln(t) -2 \right)
             \ln\left({4 \omega^2 \over m^2}\right)
\nonumber \\ [2ex]
&&            +{4 m^2 -2 q^2 \over \sqrq} \Bigg\{{9 \over 2} \ln^2(t)
            -4 \ln(t) \ln(t+1) -2 \ln(t) \ln(t-1)
\nonumber \\ [2ex]
&&            +2 {\cal L}i_2\left({1 \over t}\right)
            +4 {\cal L}i_2\left(-{1 \over t}\right) \Bigg\}
          -2 + {16 m^2 -5 q^2 \over \sqrq} \ln(t) \Bigg]\,.
\end{eqnarray}
We next present the results of the hard gluon contribution to the 
partonic structure functions indicated by ${\cal F}_{i,q}^{\rm HARD}$
in Eq. (\ref{eq:4.12}). 
The results are expressed in terms of the centre of mass energy squared of the 
incoming quark and 
vector boson in Eq. (\ref{eq:4.11}) which is denoted by $\hat s=(p+q)^2$.  
The expressions for the partonic structure functions are written in
such a way that they all start with a common factor $\smq^2/q^2$ which is 
cancelled if the integration variable $z$ in Eq. (\ref{eq:4.12}) is
replaced by $z=-q^2/\smq$ i.e.
\begin{eqnarray}
\label{eq:B10}
\int_0^{z_{max}}dz\,\hat{\cal F}_{i,q}^{\rm HARD}(z,q^2,m^2)=
\int_{m^2+2m\omega}^{\infty}d\hat s\,\frac{-q^2}{\smq^2}\,
\hat{\cal F}_{i,q}^{\rm HARD} (\hat s,q^2,m^2)\,.
\end{eqnarray}
Further we introduce the shorthand notation
\begin{eqnarray}
\label{eq:B11}
\xi= {\hat s + m^2-q^2  - \sqrt{\lam} \over  \hat s + m^2 -q^2 + \sqrt{\lam} }
 \,,\quad
\mbox{with} \quad \lam= \smq^2-4 m^2 q^2\,.
\end{eqnarray}
The results for ${\cal F}_{i,q}^{\rm HARD}$ are given by
\begin{eqnarray}
\label{eq:B12}
\hat {\cal F}_{1,q}^{\rm HARD}\!\!\!&=&\!\!\! {\it v}_q^{V_1} {\it v}_q^{V_2} 
C_F {\smq^2 \over q^2} \Bigg[ \Bigg \{ (4 m^2-2 q^2){1 \over \lsq \smm} 
        +\Bigg (-2 m^4+{7 m^2 q^2 \over 2} 
\nonumber \\[2ex]
&&       -q^4\Bigg) {1 \over \lcu}
        +\left (2 m^2 + {q^2 \over 2}\right) {\hat s \over \lcu}
        +\left (2 m^2 q^2 -{q^4 \over 2} \right) {1 \over \lsq \smq^2}
\nonumber \\[2ex]
&&      +\left (-6 m^2 +{q^2 \over 2} \right) {1 \over \lsq \smq} \Bigg \} \lnx 
        +\Bigg(-{2 q^6 \over \mpq^3} 
\nonumber \\[2ex]
&&     + {3 q^4 \over \mpq^2} -{5 q^2 \over 4 \mpq} 
       +{q^2 \over 4 \mmq} \Bigg) {1 \over \hat s}
       +\Bigg(-{q^6 \over \mpq^2}
\nonumber \\[2ex]
&&     +{3 q^4 \over 2 \mpq} -{q^2 \over 2} \Bigg) {1 \over \hat s^2}
       +\Bigg(-2 m^2 +{q^4 \over 4 m^2} +{q^4 \over \mmq}
       +{q^2 \over 2} \Bigg) {1 \over \lam}
\nonumber \\[2ex]
&&   +\Bigg(2-{q^2 \over 4 m^2} -{q^2 \over 4 \mmq} \Bigg) {\hat s \over \lam}
     +{4 \over \smm}+\Bigg(-{q^6 \over \mpq^2}
\nonumber \\[2ex]
&&     +{3 q^4 \over 2 \mpq} +2 q^2 \Bigg) {1 \over \smq^2} + \Bigg(-6
       +{q^2 \over 4 m^2} +{2 q^6 \over \mpq^3} 
\nonumber \\[2ex]
&&     -{3 q^4 \over \mpq^2} +{5 q^2 \over 4 \mpq} \Bigg) 
       {1 \over \smq} \Bigg]
\nonumber \\[2ex]
&&     + {\it a}_q^{V_1} {\it a}_q^{V_2}C_F
       {\smq^2 \over q^2}
       \Bigg[ \Bigg \{
       \Bigg(\!\!-{16 m^4 \over q^2} + 12 m^2 -2 q^2 \Bigg) {1 \over \lsq \smm}
\nonumber \\[2ex]
&&       +\Bigg(-4 m^4\!\! +\!{7 m^2 q^2 \over 2}
       -q^4 \Bigg) {1 \over \lcu}
       +\Bigg(4 m^2 +{q^2 \over 2} \Bigg) {\hat s \over \lcu}
       -\Bigg (16 m^4+{q^4 \over 2}\Bigg)
\nonumber \\[2ex]
&&        \times {1 \over \lsq \smq^2}
       +({16 m^4 \over q^2} -16 m^2 +{q^2 \over 2} \Bigg) {1 \over \lsq \smq}
       \Bigg \} \lnx
\nonumber \\[2ex]
&&       +\Bigg(-{2 q^6 \over \mpq^3} 
     +{5 q^4 \over \mpq^2} -{17 q^2 \over 4 \mpq} -{3 q^2 \over 4 \mmq} \Bigg) 
       {1 \over \hat s} 
\nonumber \\[2ex]
&&     + \Bigg(-{q^6 \over \mpq^2} 
       +{3 q^4 \over 2 \mpq} - {q^2 \over 2} \Bigg) {1 \over \hat s^2}
       +\Bigg(-4 m^2 + {q^4 \over 4 m^2} -{3 q^4 \over \mmq}
\nonumber \\[2ex]
&&   -{9 q^2 \over 2} \Bigg) {1 \over \lam} +\Bigg( 4 -{q^2 \over 4 m^2}
       +{3 q^2 \over 4 \mmq} \Bigg) {\hat s \over \lam}
       +\Bigg(4 -{16 m^2 \over q^2} \Bigg) {1 \over \smm}
\nonumber \\[2ex]
&&       +\Bigg(-16 m^2 
       -{q^6 \over \mpq^2 }+{7 q^4 \over 2 \mpq} \Bigg) {1 \over \smq^2}
       +\Bigg(-8 +{16 m^2 \over q^2}
\nonumber \\[2ex]
&&     +{q^2 \over 4 m^2}
      +{2 q^6 \over \mpq^3} -{5 q^4 \over \mpq^2}
       +{17 q^2 \over 4 \mpq} \Bigg) {1 \over \smq} \Bigg] \,,
\end{eqnarray}
\begin{eqnarray}
\label{eq:B13}
\hat {\cal F}_{2,q}^{\rm HARD}\!\!\!&=&\!\!\! {\it v}_q^{V_1} {\it v}_q^{V_2}
 C_F {\smq^2 \over q^2} \Bigg[ \Bigg \{
        (8 m^2 -4 q^2) {1 \over \lsq \smm} 
       +\Bigg(3 m^2 q^2 +{q^6 \over 4 m^2} 
\nonumber \\[2ex]
&&       -{27 q^4 \over 4} \Bigg) {1 \over \lcu}
       +\Bigg(-12 m^4 q^4 +21 m^2 q^6 -3 q^8 \Bigg) {1 \over \lfi}
        +\Bigg(-{q^4 \over 4 m^2} +5 q^2 \Bigg) {\hat s \over \lcu}
\nonumber \\[2ex]
&&       +(-36 m^2 q^4 +3 q^6) {\hat s \over \lfi}
         +\Bigg(-8 m^2 +{q^4 \over 4 m^2} -q^2 \Bigg) {1 \over \lsq \smq}
         \Bigg \} \lnx
\nonumber \\[2ex]
&&       +\Bigg({q^6 \over \mmq^3} -{q^4 \over 2 \mpq^2}
              -{q^4 \over \mmq^2}
       +{3 q^2 \over 4 \mpq}
\nonumber \\[2ex]
&&        -{3 q^2 \over 4 \mmq} \Bigg) {1 \over \hat s}
       +\Bigg(-{q^4 \over 2 \mpq} -{q^4 \over 2 \mmq} \Bigg) {1 \over \hat s^2}
\nonumber \\[2ex]
&&       +\Bigg(-{q^4 \over 4 m^2} + {4 q^8 \over \mmq^3}-{2 q^6 \over \mmq^2}
              -{9 q^4 \over 2 \mmq} +7q^2 \Bigg) {1 \over \lam}
\nonumber \\[2ex]
&&     +\Bigg(-36 m^2 q^4 +{12 q^8 \over \mmq} +27 q^6 \Bigg) {1 \over \lam^2}
       +\Bigg({q^2 \over 4 m^2} -{q^6 \over \mmq^3} 
\nonumber \\[2ex]
&&       +{q^4 \over \mmq^2} 
              +{3 q^2 \over 4 \mmq}\Bigg) {\hat s \over \lam}
       +\Bigg(-{3 q^6 \over \mmq}-12 q^4\Bigg) {\hat s \over \lam^2}
       +{8 \over \smm}
\nonumber \\[2ex]
&&       +\Bigg(-8 -{q^2 \over 4 m^2} + {q^4 \over 2 \mpq^2} 
         -{3 q^2 \over 4 \mpq} \Bigg) {1 \over \smq}  \Bigg]
\nonumber \\[2ex]
&&      +{\it a}_q^{V_1} {\it a}_q^{V_2} C_F
        {\smq^2 \over q^2}
        \Bigg[ \Bigg \{
        (8 m^2 -4 q^2) {1 \over \lsq \smm} 
       -\Bigg(6 m^2 q^2 -{q^6 \over 4 m^2} 
\nonumber \\[2ex]
&&        +{31 q^4 \over 4} \Bigg) {1 \over \lcu}
       +\Bigg(-48 m^4 q^4 +33 m^2 q^6 -3 q^8 \Bigg) {1 \over \lfi}
       +\Bigg(-{q^4 \over 4 m^2} +6 q^2 \Bigg) {\hat s \over \lcu}
\nonumber \\[2ex]
&&       +(-48 m^2 q^4 +3 q^6) {\hat s \over \lfi}
      +\Bigg(-8 m^2 +{q^4 \over 4 m^2} -2 q^2 \Bigg) {1 \over \lsq \smq}
       \Bigg\} \lnx
\nonumber \\[2ex]
&&       +\Bigg(-{3q^6 \over \mmq^3} -{5q^4 \over 2 \mpq^2}
              -{9 q^4 \over \mmq^2} + {7 q^2 \over 4 \mpq}
\nonumber \\[2ex]
&& -{23 q^2 \over 4 \mmq} \Bigg) {1 \over \hat s} +\Bigg(-{5 q^4 \over 2 \mpq} 
+{3 q^4 \over 2 \mmq}+4 q^2 \Bigg) {1 \over \hat s^2}
\nonumber \\[2ex]
&&       +\Bigg(-{q^4 \over 4 m^2} - {12 q^8 \over \mmq^3}-{42 q^6 \over \mmq^2}
              -{85 q^4 \over 2 \mmq} -6 q^2 \Bigg) {1 \over \lam}
\nonumber \\[2ex]
&&    +\Bigg(-96 m^2 q^4 -{36 q^8 \over \mmq} -21 q^6 \Bigg) {1 \over \lam^2}
       +\Bigg({q^2 \over 4 m^2} +{3 q^6 \over \mmq^3} 
\nonumber \\[2ex]
&&   +{9 q^4 \over  \mmq^2} +{23 q^2 \over 4 \mmq} \Bigg){\hat s \over \lam}
     +{9 q^6 \over \mmq} {\hat s \over \lam^2} +{8 \over \smm}
\nonumber \\[2ex]
&&       +\Bigg(-8 -{q^2 \over 4 m^2} + {5 q^4 \over 2 \mpq^2} 
       -{7 q^2 \over 4 \mpq} \Bigg) {1 \over \smq}  \Bigg]\,,
\end{eqnarray}
\begin{eqnarray}
\label{eq:B14}
\hat {\cal F}_{3,q}^{\rm HARD}\!\!\!\!\!&=&\!\!\! \Big ({\it v}_q^{V_1} 
{\it a}_q^{V_2}+{\it a}_q^{V_1} {\it v}_q^{V_2}\Big ) C_F
        {\smq^2 \over 2\,q^2}
        \Bigg[ \Bigg \{
        (16 m^2 -8 q^2) {1 \over \lsq \smm}
\nonumber \\[2ex]
&&   +(12 m^2 q^2 -8 q^4) {1 \over \lcu} +4 q^2 {\hat s \over \lcu}
       -{16 m^2  \over \lsq \smq} \Bigg \} \lnx
       +\Bigg(-{2 q^4 \over \mpq^2}  
\nonumber \\[2ex]
&&   +{2 q^2 \over \mpq}  +{2 q^2 \over \mmq}\Bigg) {1 \over \hat s}
       +\Bigg(-{2 q^4 \over \mpq } +2 q^2 \Bigg) {1 \over \hat s^2}
       +\Bigg({2 q^4 \over m^2} 
\nonumber \\[2ex]
&&   +{8 q^4 \over \mmq}  +20 q^2 \Bigg) {1 \over \lam}
     +\Bigg(-{2 q^2 \over m^2 }-{2 q^2 \over \mmq} \Bigg) {\hat s \over \lam}
       +{16 \over \smm} 
\nonumber \\[2ex]
&&       +\Bigg( -16 +{2 q^2 \over m^2} +{2 q^4 \over \mpq^2}
              -{2 q^2 \over \mpq} \Bigg) {1 \over \smq}
       \Bigg]\,.
\end{eqnarray}
The computation of the polarized structure functions proceeds in a similar way.
We obtain
\begin{eqnarray}
\label{eq:B15}
\hat g_{1,q}^{\rm HARD}\!\!\!&=&\!\!\!{\it v}_q^{V_1} {\it v}_q^{V_2} C_F 
{\smq^2 \over q^2} 
          \Bigg[ \Bigg \{
         (4 m^2 -2 q^2){1 \over \lsq \smm}
        +\Bigg({3 m^2 q^2 \over 4} -{9 q^4 \over 4}\Bigg) {1 \over \lcu}
\nonumber \\[2ex]
&&        +(3 m^4 q^4+3 m^2 q^6) {1 \over \lfi}
        +{5 q^2 \over 4} {\hat s \over \lcu}
        -15 m^2 q^4 {\hat s \over \lfi}
\nonumber \\[2ex]
&&        +\Bigg(-4 m^2 -{q^2 \over 4} \Bigg) {1 \over \lsq \smq} 
         \Bigg \} \lnx
       +\Bigg({q^6 \over \mmq^3}+{q^4 \over \mmq^2}
\nonumber \\[2ex]
&&              -{q^2 \over 4 \mpq}+{q^2 \over 4 \mmq} \Bigg) {1 \over \hat s}
       +\Bigg(-{q^4 \over 2 \mmq} -{q^2 \over 2} \Bigg) {1 \over \hat s^2}
       +\Bigg(-{q^4 \over 2 m^2}
\nonumber \\[2ex]
&&       +{4 q^8 \over \mmq^3}+{6 q^6 \over \mmq^2}
              +{7 q^4 \over 2 \mmq} +{7 q^2 \over 4} \Bigg) {1 \over \lam}
\nonumber \\[2ex]
&&       +\Bigg(-3 m^2 q^4 +{12 q^8 \over \mmq} 
       +18 q^6 \Bigg) {1 \over \lam^2}
       +\Bigg({q^2 \over 2 m^2} -{q^6 \over \mmq^3}
\nonumber \\[2ex]
&&        -{q^4 \over \mmq^2} -{q^2 \over 4 \mmq} \Bigg) {\hat s \over \lam}
       +\Bigg(-{3 q^6 \over \mmq}-9 q^4\Bigg) {\hat s \over \lam^2}
       +{4 \over \smm} 
\nonumber \\[2ex]
&&       +\Bigg(-4 - {q^2 \over 2 m^2} 
        +{q^2 \over 4 \mpq} \Bigg) {1 \over \smq}
         \Bigg]
\nonumber \\[2ex]
&&      +{\it a}_q^{V_1} {\it a}_q^{V_2} C_F {\smq^2 \over q^2} 
          \Bigg[ \Bigg \{
         (4 m^2 -2 q^2){1 \over \lsq \smm}
        +\Bigg(4 m^4-{13 m^2 q^2 \over 4} 
\nonumber \\[2ex]
&&        -{9 q^4 \over 4}\Bigg) {1 \over \lcu}
        +(-12 m^6 q^2+3 m^4 q^4+3 m^2 q^6) {1 \over \lfi}
        +\Bigg(-4 m^2 +{5 q^2 \over 4} \Bigg){\hat s \over \lcu}
\nonumber \\[2ex]
&&        +\Bigg(12 m^4 q^2-15 m^2 q^4 \Bigg){\hat s \over \lfi}
        -{q^2 \over 4} {1 \over \lsq \smq} 
         \Bigg \} \lnx
\nonumber \\[2ex]
&&       +\Bigg(-{q^6 \over \mmq^3}-{q^4 \over \mpq^2}
              -{4 q^4 \over \mmq^2}
              -{q^2 \over 4 \mpq}
\nonumber \\[2ex]
&&       -{15 q^2 \over 4 \mmq} \Bigg) {1 \over \hat s}
       +\Bigg(-{q^4 \over  \mpq} +{q^4 \over 2 \mmq}
              +{3 q^2 \over 2} \Bigg) {1 \over \hat s^2}
\nonumber \\[2ex]
&&  +\Bigg(4 m^2 -{q^4 \over 2 m^2}-{4 q^8 \over \mmq^3}-{18 q^6 \over \mmq^2}
              -{47 q^4 \over 2 \mmq} -{17 q^2 \over 4} \Bigg) {1 \over \lam}
\nonumber \\[2ex]
&&      -\Bigg(12 m^4 q^2+33 m^2 q^4 +{12 q^8 \over \mmq} +6 q^6 \Bigg) 
{1 \over \lam^2} -\Bigg(4-{q^2 \over 2 m^2} -{q^6 \over \mmq^3}
\nonumber \\[2ex]
&&    -{4 q^4 \over \mmq^2} -{15 q^2 \over 4 \mmq} \Bigg) {\hat s \over \lam}
       +\Bigg(12 m^2 q^2+{3 q^6 \over \mmq}-3 q^4\Bigg) {\hat s \over \lam^2}
\nonumber \\[2ex]
&&        +\Bigg(-{q^2 \over 2 m^2} +{q^4 \over \mpq^2}
+{q^2 \over 4 \mpq} \Bigg) {1 \over \smq}
\nonumber \\[2ex]
&& +{4 \over \smm}    \Bigg]\,,
\end{eqnarray}
\begin{eqnarray}
\label{eq:B16}
\hat g_{2,q}^{\rm HARD}\!\!\!&=&\!\!\!{\it v}_q^{V_1} {\it v}_q^{V_2} C_F 
{\smq^2 \over q^2} \Bigg[ \Bigg \{
          (-m^2 q^2 +2 q^4) {1 \over \lcu}
         -(3 m^4 q^4 +3 m^2 q^6) {1 \over \lfi}
\nonumber \\[2ex]
&&         +15 m^2 q^4 {\hat s \over \lfi}     
         \Bigg \} \lnx
         -\Bigg({q^6 \over \mmq^3} + {q^4 \over \mmq^2} +
              {q^2 \over 2 \mmq}\Bigg) {1 \over \hat s} 
\nonumber \\[2ex]
&&         +\Bigg({q^4 \over 2 \mmq} +{q^2 \over 2} \Bigg) 
                 {1 \over \hat s^2}
         -\Bigg( {4 q^8 \over \mmq^3} +{6 q^6 \over \mmq^2}
                 +{9 q^4 \over 2 \mmq} 
\nonumber \\[2ex]
&&               +{3 q^2 \over 2 } \Bigg) {1 \over \lam}
         +\Bigg( 3 m^2 q^4 -{12 q^8 \over \mmq } -18 q^6\Bigg) 
               {1 \over \lam^2}
         +\Bigg({q^6 \over \mmq^3} 
\nonumber \\[2ex]
&&           +{q^4 \over \mmq^2}
                +{q^2 \over 2 \mmq} \Bigg) 
               {\hat s \over \lam}
         +\Bigg({3 q^6 \over \mmq} +9 q^4 \Bigg) {\hat s \over \lam^2}
           \Bigg]
\nonumber \\[2ex]
&& \!\!\! + {\it a}_q^{V_1} {\it a}_q^{V_2} C_F {\smq^2 \over q^2} 
          \Bigg[ \Bigg \{
          (-m^2 q^2 +2 q^4) {1 \over \lcu}
         +(12 m^6 q^2 -3 m^4 q^4 
\nonumber \\[2ex] 
&&         -3 m^2 q^6) {1 \over \lfi}
         +(-12 m^4 q^2+15 m^2 q^4 ){\hat s \over \lfi}    
         \Bigg \} \lnx
         +\Bigg({q^6 \over \mmq^3} + {4q^4 \over \mmq^2} 
\nonumber \\[2ex]
&&               + {7 q^2 \over 2 \mmq}\Bigg) {1 \over \hat s} 
         -\Bigg({q^4 \over 2 \mmq} +{q^2 \over 2} \Bigg) 
                 {1 \over \hat s^2}
         +\Bigg({4 q^8 \over \mmq^3} 
\nonumber \\[2ex]
&&     +{18 q^6 \over \mmq^2} +{45 q^4 \over 2 \mmq} +{17 q^2 \over 2 }
               \Bigg) {1 \over \lam}
         +\Bigg(12 m^4 q^2 +33 m^2 q^4 
\nonumber \\[2ex]
&&         +{12 q^8 \over \mmq } +6 q^6 \Bigg) {1 \over \lam^2}
         -\Bigg({q^6 \over \mmq^3} +{4 q^4 \over \mmq^2}
                +{7 q^2 \over 2 \mmq} \Bigg) 
               {\hat s \over \lam}
\nonumber \\[2ex]
&&         -\Bigg(12 m^2 q^2 +{3 q^6 \over \mmq} -3 q^4 \Bigg) 
{\hat s \over \lam^2} \Bigg]\,,
\end{eqnarray}
\begin{eqnarray}
\label{eq:B17}
\hat g_{3,q}^{\rm HARD}\!\!\!&=&\!\!\! \Big ({\it v}_q^{V_1} {\it a}_q^{V_2}
+{\it a}_q^{V_1} {\it v}_q^{V_2}\Big ) C_F 
{\smq^2 \over 2\,q^2} 
          \Bigg[ \Bigg \{
         (-16 m^2 +8 q^2) {1 \over \lsq \smm}
\nonumber \\[2ex]
&&     +\Bigg(m^2 q^2 +{q^6 \over m^2}   +22 q^4 \Bigg) {1 \over \lcu}
        +\Big (12 m^2 q^6 -12 q^8\Big ) {1 \over \lfi}
        +\Bigg (-{q^4 \over m^2} -13 q^2\Bigg) {\hat s \over \lcu}
\nonumber \\[2ex]
&&       +\Big (96 m^2 q^4 +12 q^6\big ) {\hat s \over \lfi}
        +\Bigg(16 m^2 +{q^4 \over m^2} +5 q^2 \Bigg) {1 \over \lsq \smq}
         \Bigg \} \lnx
\nonumber \\[2ex]
&&       +\Bigg({2 q^4 \over \mpq^2}  +{2 q^4 \over \mmq^2}
         +{2 q^2 \over \mmq} \Bigg) {1 \over \hat s}
        +\Bigg({2 q^4 \over \mpq } -2 q^2 \Bigg) {1 \over \hat s^2}
\nonumber \\[2ex]
&&        +\Bigg({5 q^4 \over m^2} +{8 q^6 \over \mmq^2 }
        +{12 q^4 \over \mmq} -5 q^2 \Bigg) {1 \over \lam}
        +(48 m^2 q^4 -24 q^6) {1 \over \lam^2}
\nonumber \\[2ex]
&&        +\Bigg(-{5 q^2 \over m^2} -{2 q^4 \over \mmq^2}
          -{2 q^2 \over \mmq} \Bigg) {\hat s \over \lam}
        +48 q^4 {\hat s \over \lam^2} 
\nonumber \\[2ex]
&&        -{16 \over \smm}
        +\Bigg (16 +{5 q^2 \over m^2} 
        -{2 q^4 \over \mpq^2} \Bigg) {1 \over \smq} \Bigg]\,,
\end{eqnarray}
\begin{eqnarray}
\label{eq:B18}
\hat g_{4,q}^{\rm HARD}\!\!\!&=&\!\!\! \Big ({\it v}_q^{V_1} {\it a}_q^{V_2}+
{\it a}_q^{V_1} {\it v}_q^{V_2}\Big ) C_F 
{\smq^2 \over 2\,q^2} 
          \Bigg[ \Bigg \{
         (-16 m^2 +8 q^2 ) {1 \over \lsq \smm}
\nonumber \\[2ex]
&&  +\Bigg(-3 m^2 q^2    + {q^6 \over m^2} +18 q^4 \Bigg) {1 \over \lcu}
        +(-108 m^4 q^4 -156 m^2 q^6 ) {1 \over \lfi}
\nonumber \\[2ex]
&& +\Bigg(-{q^4 \over m^2} -13 q^2\Bigg) {\hat s \over \lcu} 
        +12 m^2 q^4 {\hat s \over \lfi}
        +\Big (-120 m^4 q^8 +120 m^2 q^{10}\Big ) {1 \over \lse}
\nonumber \\[2ex]
&&  +\Big (-960 m^4 q^6 -120 m^2 q^8\Big ) {\hat s \over \lse} +\Bigg (16 m^2 
        +{q^4 \over m^2} +5 q^2\Bigg ) 
\nonumber \\[2ex]
&& \times {1 \over \lsq \smq} \Bigg \} \lnx
 +\Bigg(-{12 q^8 \over \mmq^4}  -{32 q^6 \over \mmq^3}
         +{2 q^4 \over \mpq^2 }
\nonumber \\[2ex]
&& -{20 q^4 \over \mmq^2} \Bigg){1 \over \hat s}+\Bigg({2 q^6 \over \mmq^2} 
 +{2 q^4 \over \mpq}  +{4 q^4 \over \mmq} \Bigg) {1 \over \hat s^2}
        +\Bigg({5 q^4 \over m^2} 
\nonumber \\[2ex]
&& -{48 q^{10} \over \mmq^4} -{152 q^8 \over \mmq^3}   
-{146 q^6 \over \mmq^2}-{44 q^4 \over \mmq} -11 q^2 \Bigg) {1 \over \lam}
\nonumber \\[2ex]
&&        +\Bigg(-156 m^2 q^4 -{80 q^{10} \over \mmq^2}
   -{248 q^8 \over \mmq}  -216 q^6 \Bigg) {1 \over \lam^2}
        +\Bigg(-{5 q^2 \over m^2} 
\nonumber \\[2ex]
&& + {12 q^8 \over \mmq^4}+{32 q^6 \over \mmq^3}  +{20 q^4 \over \mmq^2}\Bigg) {\hat s \over \lam} 
      +\Bigg({20 q^8 \over \mmq^2} + {52 q^6 \over \mmq} 
\nonumber \\[2ex]
&&  +60 q^4 \Bigg){\hat s \over \lam^2}  
-{16 \over \smm}  +\Big (-480 m^4 q^6+240 m^2 q^8\Big ) {1 \over \lam^3}
        -480 m^2 q^6 {\hat s \over \lam^3}
\nonumber \\[2ex]
&& +\Bigg(16 +{5 q^2 \over m^2}-{2 q^4 \over \mpq^2} 
\Bigg) {1 \over \smq} \Bigg]\,,
\end{eqnarray}
\begin{eqnarray}
\label{eq:B20}
\hat g_{5,q}^{\rm HARD}\!\!\!&=&\!\!\! \Big ({\it v}_q^{V_1} {\it a}_q^{V_2}
+{\it a}_q^{V_1} {\it v}_q^{V_2}\Big ) C_F 
{\smq^2 \over 2\,q^2} 
          \Bigg[ \Bigg \{
         (-8 m^2 +4 q^2) {1 \over \lsq \smm}
        +\Bigg({15 m^2 q^2 \over 2} 
\nonumber \\[2ex]
&&        + {11 q^4 \over 2} \Bigg) {1\over \lcu}
        +(-24 m^6 q^2 +42 m^4 q^4 -18 m^2 q^6) {1 \over \lfi}
        -{7 q^2 \over 2} {\hat s \over \lcu}
        +(24 m^4 q^2 
\nonumber \\[2ex]
&&        +30 m^2 q^4) {\hat s \over \lfi}
        +\Bigg(8 m^2 +{3 q^2 \over 2} \Bigg) {1 \over \lsq \smq}
        \Bigg \} \lnx
        +\Bigg({q^4 \over \mpq^2} 
\nonumber \\[2ex]
&&         - {q^4 \over \mmq^2}
         -{3 q^2 \over 2 \mpq} -{5 q^2 \over 2 \mmq} \Bigg)
        {1 \over \hat s}
        +\Bigg({q^4 \over \mpq} -q^2 \Bigg) {1 \over \hat s^2}
\nonumber \\[2ex]
&&        +\Bigg(-{4 q^6 \over \mmq^2} -{12 q^4 \over \mmq} -{21 q^2 \over 2}
         \Bigg) {1 \over \lam}
        +(-24 m^4 q^2 +18 m^2 q^4 
\nonumber \\[2ex]
&&        -6 q^6) {1 \over \lam^2} +\Bigg({q^4 \over \mmq^2}
          +{5 q^2 \over 2 \mmq} \Bigg) {\hat s \over \lam}
        +(24 m^2 q^2 +6 q^4){\hat s \over \lam^2} 
\nonumber \\[2ex]
&&        -{8 \over \smm}
        +\Bigg(8 -{q^4 \over \mpq^2} +{3 q^2 \over 2 \mpq}\Bigg) {1 \over \smq}
        \Bigg]\,.
\end{eqnarray}


%


\end{document}